\documentclass[12pt]{article}

\usepackage[english]{babel}
\usepackage[utf8]{inputenc}
\usepackage{amsmath}
\usepackage{empheq}
\usepackage{amssymb}
\usepackage{graphicx} 
\usepackage[font=small,labelfont=bf, width=0.95\textwidth]{caption}

\usepackage{mdframed} 
\usepackage{framed} 
\usepackage[most]{tcolorbox} 
\tcbuselibrary{breakable}

\usepackage{booktabs} 

\usepackage{cancel} 

\usepackage{natbib}
\bibpunct{(}{)}{;}{a}{,}{;}

\usepackage{amsthm} 
\newtheorem{definition}{Definition}
\newtheorem{assumption}{}

\title{Pseudo-biodiversity effects across scales}
\author{ \normalsize{Pradeep Pillai$^{1,\dagger}$}\\
\normalsize{$^{1}$Marine Science Center, Northeastern University,}\\
\normalsize{430 Nahant Rd, Nahant, MA 01908}\\
\normalsize{$^{\dagger}$Corresponding author: pradeep.research@gmail.com} }

\date{}

\begin{document}
{\centering

\textbf{\Large{Pseudo-biodiversity effects across scales}} \\
\bigskip
Pradeep Pillai$^{1,\dagger}$ \\
\textit{
$^{1}$Marine Science Center, Northeastern University,\\
430 Nahant Rd, Nahant, MA 01908\\
$^{\dagger}$To whom correspondence should be addressed. E-mail: pradeep.research@gmail.com \\
}
}

\begin{abstract}
  \noindent Over the last decade several attempts have been made to extend biodiversity studies in ways that would allow researchers to explore how biodiversity-ecosystem functioning relationships may change across different spatial and temporal scales. Unfortunately, the studies based on these attempts often overlooked the serious issues that can arise when quantifying biodiversity effects at larger scales, specifically the fact that biodiversity effects measured across space and time can contain trivial effects that are unrelated to the role of biodiversity per se -- or even effects that are non-biological in nature due to being simple artefacts of how properties and entities are counted and quantified. Here we outline and describe three such pseudo-biodiversity effects: \emph{Population-level effects}, \emph{Independence effects}, and \emph{Arithmetic effects}. Population-level effects are those related to temporal changes due to \emph{individual} species population growth or development, and are thus independent of biodiversity. Independence and Arithmetic effects (which we explore here primarily in a spatial context) arise either as a simple consequence of the fact that not all species are present everywhere -- i.e., species turnover is inevitable at greater spatial scales (Independence effects); or they arise when the purported biodiversity effects measured are a simple byproduct of how mathematical functions behave (Arithmetic effects). Our study demonstrates the necessity of controlling for these trivial artefactual effects if one wishes to meaningfully measure how true biodiversity effects change across spatial and temporal scales.
\end{abstract}


\section{Introduction}
The hope of uncovering new directions in biodiversity-ecosystem functioning (BEF) research has led investigators to explore how the relationships between biodiversity and various ecosystem properties may unfold across different spatial and temporal scales. This research direction has intensified in recent years with the rise of working group review and opinion pieces \citep{gonzalez.germain.ea_20}, and with the application of meta-analyses on existing data sets \citep{qiu.cardinale_20}, supplemented further by recent attempts at experimental \citep{huang.chen.ea_18,reich2012} as well as simulation-based \citep{thompson.isbell.ea_18} work.

However, a potential danger with these new lines of enquiry is the speed with which new studies are being generated -- well before any attempts have been made to consider or address serious conceptual and logical issues that can arise as one extends biodiversity studies across larger temporal and spatial scales.  This is most strikingly apparent in the inability of researchers to control for either the trivial or confounding effects that arise as one measures the ecosystem properties associated with species while moving across time and space -- as for example, when the changes in ecosystem effects that are measured across scales are not proper \emph{biodiversity} effects arising from species being in communities, but are effects that would otherwise be observed even when species exist alone in isolation. Another, more problematic example is when the effects quantified at different scales are not even biological in nature, but instead are mere artefacts of how quantification changes when certain measures are applied at different spatial extents. Many of the spurious forms of quantification we explore here may ultimately be due -- at least in part -- to the lack of rigor when it comes to defining what exactly biodiversity is in a given context, as well as what constitutes a \emph{meaningful} change in biodiversity effects. 

Here we highlight three possible pseudo-biodiversity effects that arise when the tools that are used to measure how increasing biodiversity affects ecosystem functioning at a given moment in time or point in space are then extended to quantify effects over larger time periods or at large spatial scales. The first pseudo effect we consider, \emph{Population-level effects}, is the quantification of temporal changes in properties associated with single-species populations (e.g., growth or development), and are thus effects that would otherwise have been measured in the absence of biodiversity. \emph{Independence effects} are those that arise due to non-interacting species operating and functioning independently of each other (largely as the result of the inevitable turnover expected given the increasing scale of the natural system considered). Because the normal functioning of independent, non interacting species are responsible for generating this effect, its measurement should not be considered ecologically meaningful, and thus not considered a true biodiversity effect. The last pseudo-biodiversity effect we consider -- \emph{Arithemetic effects} -- represents quantities that are simply artefacts of how mathematical functions behave, and thus have no biological, let alone ecological meaning, whatsoever. Arithmetic effects represent an extreme case where the inevitable properties of logical rules or mathematical functions are misconstrued as natural or empirical observations. 

To demonstrate the operation of the above pseudo effects we will use in this study the two most common methods for measuring biodiversity-ecosystem functioning relationships: measures based on the \emph{total} raw biodiversity effects, and those based on the Loreau-Hector (LH) \emph{net} biodiversity effect. The total biodiversity effect, $y_{_\mathrm{Total}}$, is simply the straightforward raw or total measure of some property (e.g., yield) arising from multiple species (or other taxonomic units) occurring together in mixtures. The Loreau-Hector net biodiversity effect \citep{loreau.hector_01a}  is a portion of this total biodiversity effect calculated relative to a baseline effect arising from species merely coexisting in mixtures neutrally \citep{pillai.gouhier_19}. This baseline is the Loreau-Hector null, $y_{_\mathrm{Null}}$, which is defined, for a mixture with $n$ species and an average monocultue yield $\overline{Y}$, as $y_{_\mathrm{Null}} =n \times \overline{Y}$. This null is subtracted from the total biodiversity effect, $y_{_\mathrm{Total}}$, to give the net effect, $y_{_\mathrm{Net}}$:
\begin{align}
  y_{_\mathrm{Net}} &= y_{_\mathrm{Total}} - y_{_\mathrm{Null}}. 
\end{align}
By attempting to control for the most trivial and default effects associated with species merely coexisting together, Loreau and Hector's net biodiversity effect represented an important improvement over the use of the raw total biodiversity effect in previous biodiversity studies. Because some ecosystem functioning will always be measured simply by species existing, whether alone or neutrally together in mixtures (i.e., sharing a habitat niche in a zero-sum manner), the value of the LH null lies in its ability to allow this default effect to be discounted. If species exist in a neutral community without interacting then the LH net effect should be zero, $y_{_\mathrm{Net}}=0$, even if the total biodiversity effect measured is positive.

In most studies using the Loreau-Hector method the net biodiversity effect is further partitioned into a complementarity effect (CE), which measures the degree that species' effects on each other's ecosystem functioning (positive and negative) are symmetrical,  and the selection effect (SE), which measures the degree to which species competitively dominate the mixture at the expense of others in a zero-sum manner,  $y_{_\mathrm{Net}} = y_{_\mathrm{CE}} +y_{_\mathrm{SE}}$ \citep{pillai.gouhier_19}.  Despite being an improvement over previous ecosystem functioning measures, the LH method suffers from a critical flaw when individual species in monocultures have a nonlinear abundance-ecosystem functioning relationship. This nonlinearity in monocultures will render the LH null incoherent and its partitioning into complementarity and selection effects completely spurious \citep{pillai.gouhier_19, pillai.gouhier_19a}.

Although these spurious biodiversity effects (CE and SE) can be amplified at both larger temporal and spatial scales, we will for the most part not consider them here, except in the special case involving spurious CE effects that are inflated over time -- the demonstration of which will then be used to explain some of the patterns observed in recent studies and meta analyses. Otherwise, when dealing with  BEF across spatial scales, the pseudo effects we explore here will not be the result of any spurious LH biodiversity effects, either because we will assume linearity, or because the LH biodiversity effect is not germane to the pseudo effect in question.    

The study we present here is an analytical demonstration of how pseudo biodiversity effects may be contained (and are otherwise undetectable) within the measurements of biodiversity effects that have been quantified at different scales. Analytically separating out these effects requires that a set of highly idealized and simplifying assumptions are made in order to elucidate how pseudo effects can be inherently confounded with true biodiversity effects in experiments. To help the reader follow the deductive arguments that follow more easily we will explicitly spell out the assumptions made for each set of demonstrations  (e.g., Box 1 with Key Assumptions). The role of the analytical work presented below then is not to model empirical systems as they are, but to demonstrate how (unbeknownst to current researchers)  measurements in biodiversity experiments are likely to confound trivial pseudo effects with real biodiversity effects. This paper thus aims to provide  an early and critical intervention for highlighting the need to address and account for these confounding issues before the research program develops any further.

\section{Pseudo-biodiversity effects across time: the role of population-level change effects}
In this section we consider how population-level effects leak into or become mischaracterized as \emph{biodiversity} effects. This pseudo-biodiversity effect is a confounding effect in studies that attempt to measure ecosystem functioning at different temporal scales  \citep[for example,][]{reich2012,huang.chen.ea_18,qiu.cardinale_20}.  We will begin with our definition of Population-level effects as follows
\begin{definition}[Population-level effects]\label{d1}
  Observed changes in ecosystem effects over time that arise at the level of a single-species population independent of other populations. Specifically, changes in ecosystem properties or effects that arise due to growth, development or other changes that a single species population undergoes independent of other, co-occurring species. In mixtures population-level effects refers to those effects that would have been measured in the absence of other species.
\end{definition}

In order to study how how a pseudo-biodiversity effect like Population-level effects operate over time, we will first consider how ecosystem-functioning -- specifically, yields in plant monocultures -- change  over time on their own without being in mixtures.  (As a side note: for a large portion of this paper, and more specifically for this section, we will assume that the ecosystem functioning we are considering is plant yield or biomass, unless otherwise stated. This allows us to directly relate the analytical arguments obtained here directly to the empirical patterns seen in the literature; needless to say, the pseudo-biodiversity effects we elucidate here, nevertheless, hold for other types of functioning.)

\begin{figure}[h]
  \centering
  \includegraphics[width=0.60\textheight]{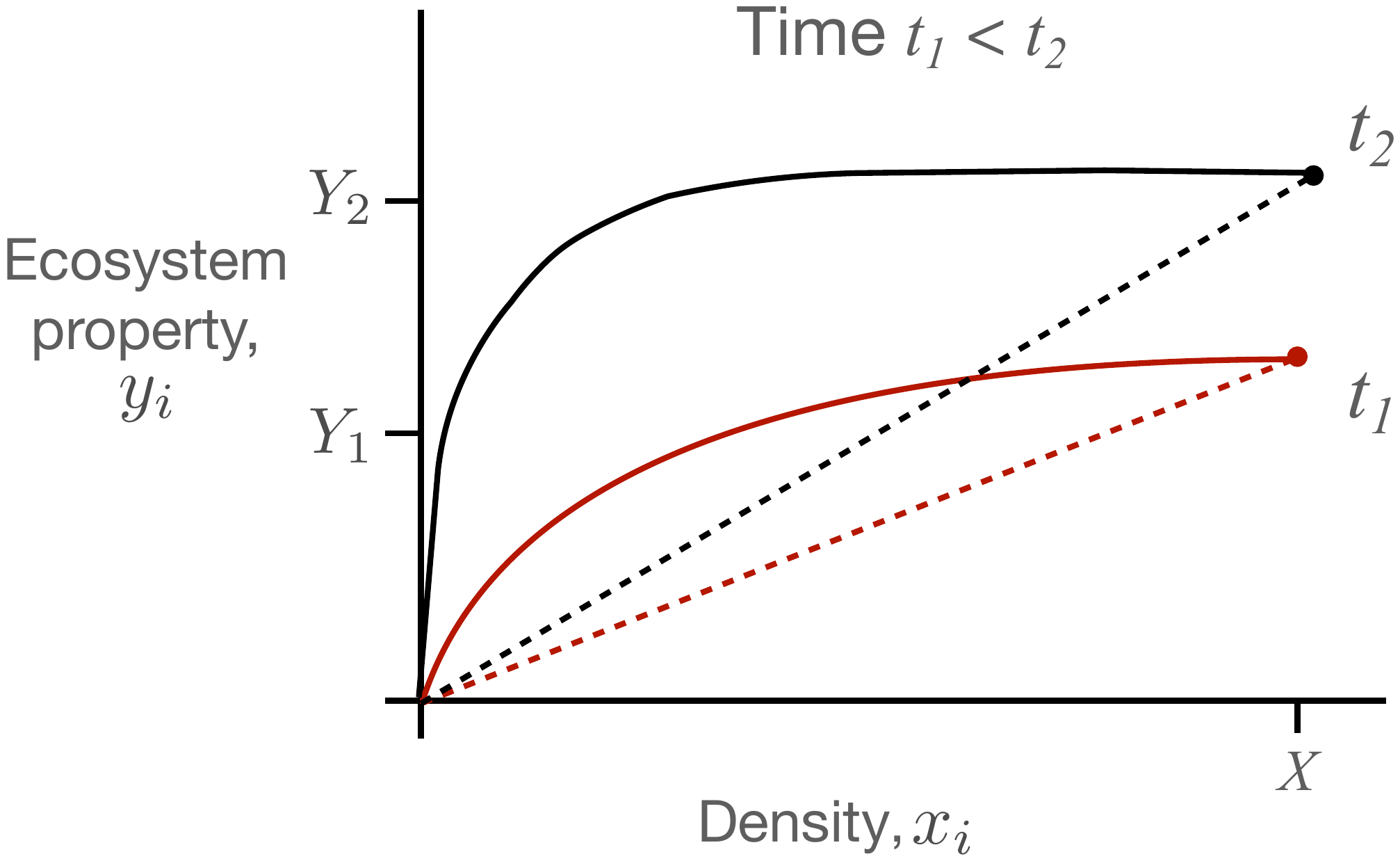}
  \caption{Shift in monoculture yields over time}{}
  \label{Fig:concave_yields}
\end{figure}

In plant or crop systems the typical pattern for changes in yield of a single species (monoculture) population over time  is for both the concavity of the  yield curve and the monoculture yield itself (measured at a given population density) to increase (Fig \ref{Fig:concave_yields}). Let us take a concave monoculture yield curve relating how yield, $y_i$, for a species $i$ changes with population density $x_i$, and where the parameter $Y_i$ represents the yield that corresponds to the maximum density $X_i$ (Fig \ref{Fig:concave_yields}). The concave function will be of the form $y_i = A_i x_i^{k_i}$, where $k_i$ is the exponent that determines the degree of concavity of this single species population ($0 \leq k_i \leq 1$), and $A_i$ is a coefficient with the value $A_i = Y_i/X^{k_i}$. The secant line between the origin and the point given by yield $Y_i$ corresponds to a linear abundance-yield curve of the form $(Y_i/X_i)\times x_i$, where $(Y_i/X_i)$ is the slope.

If one only knows the relationship between the final yield, $Y_i$, corresponding to maximum density $X_i$, and one wished to predict the yield that would be obtained when the population was only at a fraction, $p$ of the maximum density, $pX_i$, then this predicted yield will depend on the functional form of the abundance-yield curve. As we will see, the degree to which the concave curve is greater than the secant line is the degree to which concavity artificially inflates the biodiversity effect in a single species, ${\Delta}_i$. Recall that the net biodiversity effect is given by the `\emph{observed yield}' $-$ `\emph{predicted (null) yield}'. Then for a given species $i$, the inflation of its contribution to the net biodiversity effect is given by
\begin{align}
  { \Delta}_i &= A_i x_i^{k_i} -  \frac{Y_i}{X_i} x_i \notag \\
  &=  (p^{k_i} -p)Y_i. 
\end{align}

In the Loreau-Hector method, the predicted or expected yields in mixtures are expressed as proportions of the final monoculture yields, and are thus ultimately based on a transformation of proportional \emph{densities} ($pX_i$) into proportional \emph{properties} or yields ($pY_i$). There is the implicit assumption that the secant line will give the true yield expectations, and it is this implicit assumption that allows the LH method to be coherent, and to work as a measurement of biodiversity effects. In the standard LH null the proportions of initial densities are  given by $p=1/n$. As a result, the expected yields for each  species $i$ in a mixture will be  $\frac{1}{n} Y_i$.

From here on in we will assume that all polycultures or mixtures are neutral communities, where there is no competitive dominance and all species equally share the habitat niche; as well, all densities are fixed at their initial proportions $p=1/n$, as predicted by the LH null. Under this assumption, without positive or negative interactions, the LH null yield for the whole community mixture should be zero.

Summing the inflated biodiversity effect of each species, ${\Delta}_i$, across all species in a mixture will give a measure of how the \emph{net} biodiversity effect for the community has been artificially inflated by concavity of the monoculture abundance-yield curve, despite an expectation of zero:
\begin{align} \label{eq:Delta_net}
  \Delta_{\mathrm{Net}} &= \sum_{i=1}^n (n^{-k_i} - n^{-1})Y_i  \notag \\
                        &= \left(n^{1-k} - 1 \right) \overline{Y}, 
\end{align}
for all diversity levels $ n \in  \{ 1,2,\ldots \infty \} $. When monoculture abundance-functioning relations for all species do not follow the secant line ($k\neq 1$), then there will be an artificial net biodiversity effect measured that is greater than zero. This effect is not due to \emph{biodiversity} itself, but arises directly from the effects of \emph{single species monocultures}. In other words, single species or \emph{population-level} effects,  are `leaking' into the measurement of the net biodiversity effects. In real experiments, these population-level effects have the potential to be confounded with true diversity effects. In the simple cases where each species has the same concave or saturating density-functioning relationship, the total inflation of the net biodiversity effect is equivalent to the inflated complementarity effect $\Delta_{\mathrm{Net}}=\Delta_{\mathrm{CE}}$.

Similarly, we can also sum across all species to get the \emph{total raw} biodiversity effect in a mixture,
\begin{align} \label{eq:total_biodiv_effect}
  y_{_{\mathrm{Total}}} &= \overline{Y} n^{1-k}.  
\end{align}
(Note that the net biodiversity effect is just the portion of this raw total biodiversity effect left over after the expected yield has been subtracted.) Throughout this paper, we will analyse how both the artificially inflated net effect, $\Delta_{\mathrm{Net}}$, and total raw biodiversity effect, $y_{_{\mathrm{Total}}}$ changes as one measures them across different spatial and temporal scales, and the degree to which the changes measured represent pseudo-biodiversity effects.

Before exploring how biodiversity effects change across scales, we can briefly check to see how the artificial net effect itself changes with increasing diversity. We can do this by using a version of Eq. \eqref{eq:Delta_net} which is continuously differentiable over the domain of the positive reals,  $ n \in \mathbb{R}^+ $, and then differentiating with respect to $n$ to give
\begin{align} \label{eq:der_inflation_wrt_n}
  \frac{\partial}{\partial n} \Delta_{\mathrm{Net}} &= \overline{Y} (1-k)\;n^{-k} \; . 
\end{align}
Since the derivative with respect to $n$ increases ($\Delta' >0$) across the entire domain, this means that the artificial inflation of the net effect will actually increase as biodiversity, $n$, increases. In other words, some positive biodiversity-ecosystem effect is built-in to the experimental design of the Loreau-Hector partitioning when it comes to the properties of species that follow self-thinning rules \citep{mrad.manzoni.ea_20,westoby_84}.

Note how this artificial inflation of the net biodiversity effect is actually not a proper biodiversity effect, but a \emph{population-level} effect associated with single-species monocultures. It is an effect that arises from the concave abundance-ecosystem functioning relationship observed in \emph{monocultures}, a population-level property without which the the inflation effect disappears (as is seen when the parameter $k$ is set to 1). These population-level effects can potentially `leak' into the measurements made in polycultures and consequently, may be misconstrued as true biodiversity effects. BEF studies have historically not accounted or controlled for these single species effects,  particularly in studies that explore how both the total and the net biodiversity effect (as well as their various components) change over time, as we shall see below.

\subsubsection*{Inflation of the artificial net biodiversity effect over time}
Exploring how the built-in positive net biodiversity effect, $\Delta_{\mathrm{Net}}$, changes over time can give us a sense of the degree to which the biodiversity effect changes measured in temporal studies may be at least partly spurious. Recall in Figure \ref{Fig:concave_yields} how over a given time span the plant monoculture yields tend to increase in value while at the same time their yield curves grow increasingly concave. That is, in plants (or similar) systems we can often see for a given time period the total yields and the concavity of the ecosystem-density relationship change as follows:
\begin{align} \label{eq:Y_wrt_t}
  \frac{\mathrm{d}\overline{Y}}{\mathrm{d}t} &> 0,  \;\;\;\;\;\;\text{(Average monoculture yields go up over given time span)}, \;\;\;\;\;
\end{align}
and
\begin{align} \label{eq:k_wrt_t}
  \frac{\mathrm{d}k}{\mathrm{d}t} &< 0,  \;\;\;\;\text{(Concavity increases, i.e., $k$ decreases, over given time span)}. \;\;\;\;
\end{align}
The question we are ultimately interested in investigating is how the effects of these two variables changing over time might be confounded with the effects of increasing biodiversity over the same period. But before we do that we need to know how the net biodiversity effect itself changes over time. 

We can measure the change or inflation of the net biodiversity effect over time as follows,

\begin{align} \label{eq:net_inflation_wrt_t}
  \frac{\mathrm{d}}{\mathrm{d}t}  \Delta_{\mathrm{Net}} &= \frac{\mathrm{d}\overline{Y}}{\mathrm{d}t} \cdot \;\left( \frac{\partial  }{\partial \overline{Y}}  \Delta_{\mathrm{Net}} \right) \; + \; \frac{\mathrm{d} k}{\mathrm{d}t} \cdot \; \left( \frac{\partial }{\partial k} \Delta_{\mathrm{Net}} \right) . 
\end{align}
We see from the above expression that the growth or decline in the inflation effect will be determined by the effects entailed by $\frac{\partial }{\partial \overline{Y}} \Delta_{\mathrm{Net}}$ and $\frac{\partial }{\partial k} \Delta_{\mathrm{Net}}$. These effects are given by
\begin{align} \label{eq:der_inflation_wrt_Y}
  \frac{\partial  }{\partial \overline{Y}} \Delta_{\mathrm{Net}} &= \left(n^{1-k} - 1 \right), 
\end{align}
and
\begin{align} \label{eq:der_inflation_wrt_k}
  \frac{\partial  }{\partial k} \Delta_{\mathrm{Net}} &= -\overline{Y}\;n^{1-k} \;\log n . 
\end{align}

From the above expressions \eqref{eq:der_inflation_wrt_Y} and \eqref{eq:der_inflation_wrt_k} we know that $\frac{\partial }{\partial \overline{Y}} \Delta_{\mathrm{Net}} > 0$ and $\frac{\partial }{\partial k} \Delta_{\mathrm{Net}} <0$ always hold. Combining this with conditions \eqref{eq:Y_wrt_t} and \eqref{eq:k_wrt_t}, we see that the two terms on the RHS of Eq. \eqref{eq:net_inflation_wrt_t} will be positive, i.e.,
\begin{align} 
  \frac{\mathrm{d}}{\mathrm{d}t}  \Delta_{\mathrm{Net}} &= \;  \Bigl( +\Bigr)\cdot \Bigl( +\Bigr) \; + \; \Bigl( -\Bigr)\cdot \Bigl( -\Bigr), \notag
\end{align}
and thus the  inflation of the net biodiversity effect should grow over the relevant time period where conditions \eqref{eq:Y_wrt_t} and \eqref{eq:k_wrt_t} hold. This is a simple default change that arises \emph{not} from biodiversity, but from changes that single species are already undergoing independently of each other.

Without controlling and accounting for the default changes expected in single species over time, the measured changes in net biodiversity effects are likely to represent changes that are actually population-level effects, as demonstrated above. These same population-level effects can also arise in measurements involving the total or raw biodiversity effect, as we demonstrate briefly below. 

\subsubsection*{Inflation of the total biodiversity effect over time}
The preceding analysis was focussed on the LH net biodiversity effect, which is an arbitrary portion of the \emph{total} or raw ecosystem property measured in a given community. However, both the increase in monoculture yields and the increasingly concave monoculture relationships that occur over time have a role in not only increasing the inflation of the the \emph{net} biodiversity effect, $\Delta_{\mathrm{Net}}$, but also in increasing the total raw biodiversity effect $y_{\mathrm{Total}}$, as well. Recall the total biodiversity effect, $y_{_{\mathrm{Total}}}$ is given by
\begin{align} \label{eq:total_biodiversity}
  y_{_{\mathrm{Total}}} &= \overline{Y}\;n^{1-k}  .
\end{align}

Taking the time derivative of Eq. \eqref{eq:total_biodiversity} above (and using the explicit expression for $\Delta_{\mathrm{Net}}$) shows how the inflation of the total biodiversity effect grows over time along with the inflation of the net biodiversity effect:
\begin{align} \label{eq:total_inflation_wrt_t}
  \frac{\;\mathrm{d} y_{_{\mathrm{Total}}} }{\mathrm{d}t}  &=   \frac{\mathrm{d}}{\mathrm{d}t}  \Delta_{\mathrm{Net}} \; + \; \frac{\mathrm{d}\overline{Y}}{\mathrm{d}t}   . 
\end{align}
If only the curvature of the monoculture yields changes over time and not the yields (i.e., $ \frac{\mathrm{d}\overline{Y}}{\mathrm{d}t} =0$) then both the artificial inflation of both the raw and net biodiversity effects should change identically.
\begin{align}
  \frac{\;\mathrm{d} y_{_{\mathrm{Total}}} }{\mathrm{d}t\;\;}  &=  \frac{\;\mathrm{d}\Delta_{\mathrm{Net}}}{\mathrm{d}t \;\;}  . 
\end{align}
What these expressions indicate is that the built-in inflation of biodiversity effects over time are not limited to studies using the Loreau-Hector net biodiversity effect;  inflation effects over time can also appear in biodiversity studies that only measure the total or raw ecosystem functioning. The temporal inflation of both types of biodiversity effects may be involved in producing the patterns seen in many current of the studies of how increases in  biodiversity will increase ecosystem functioning across time, as we will demonstrate next.

\subsection{Patterns of Net/Complementarity Effect inflation with increasing biodiversity, $n$}
Recall that we assumed in our simple idealized scenario that all species have identical values for the exponent parameter $k$, and that all species in a mixture divide up the total community abundance proportionally based on their initial fraction of the total of all monoculture abundances. Under these simple assumptions the inflation of the net biodiversity effect, $\Delta_{\mathrm{Net}}$ in Eq \eqref{eq:total_biodiversity} will arise solely due to inflation of the complementarity effect, CE, that is, $\Delta_{\mathrm{Net}}=\Delta_{\mathrm{CE}}$. If the value of the parameter $k$ varies amongst species then the inflation of the net biodiversity effect will also include a component arising from inflationary selection effects (unless all species have identical monoculture effects, $Y$). 

Since we will continue to assume that all species have identical values for their $k$ exponents, our analysis of how the artificial inflation of the net biodiversity effect changes over time will be identical with an analysis of artificial inflation of complementarity effects. As such, the artificial inflation of the complementarity effect (CE) over time will exhibit the default pattern seen above, where Population-level effects become mischaracterized or leak into biodiversity effects, even when an individual species exists on its own.  

How an increase in biodiversity, $n$, will affect the artificially inflated complementarity effect over time, $\frac{\mathrm{d} }{\mathrm{d} t}  \Delta_{\mathrm{CE}}$,  will be given by the expression
\begin{align} \label{eq:biodiv_effect_der_inflation_wrt_t}
  \frac{\partial}{\partial n} \left( \frac{\mathrm{d} }{\mathrm{d} t}  \Delta_{\mathrm{CE}} \right) .
\end{align}
(Note that the expression for $\frac{\mathrm{d} }{\mathrm{d} t}  \Delta_{\mathrm{CE}}$ is identical to that for $\frac{\mathrm{d} }{\mathrm{d} t}  \Delta_{\mathrm{Net}}$ in Eq. \eqref{eq:net_inflation_wrt_t} given that $\Delta_{\mathrm{Net}}=\Delta_{\mathrm{CE}}$.)  

By inserting Eq. \eqref{eq:net_inflation_wrt_t} into the above expression we can describe the effect of increasing diversity on the artificial inflation effect over time as a sum involving the mixed partial derivatives of $\Delta_{\mathrm{CE}}$,
\begin{align} \label{eq:mixed_der_inflation_wrt_n}
  \frac{\partial}{\partial n} \left( \frac{\mathrm{d} }{\mathrm{d} t}  \Delta_{\mathrm{CE}} \right) &= \frac{\mathrm{d} k}{\mathrm{d} t} \cdot \left( \frac{\partial^2}{\partial n \;\partial k} \Delta_{\mathrm{CE}} \right)  \; + \; \frac{\mathrm{d} \overline{Y} }{\mathrm{d} t} \cdot \left( \frac{\partial^2}{\partial n \;\partial \overline{Y}} \Delta_{\mathrm{CE}} \right). 
\end{align}
After first noting that the expression for $\frac{\partial}{\partial n}  \Delta_{\mathrm{CE}}$ is identical to the expression for  $\frac{\partial}{\partial n}  \Delta_{\mathrm{Net}}$ that we saw earlier in Eq. \eqref{eq:der_inflation_wrt_n}, we can then write out the expressions for both of the mixed partial derivatives on the RHS of the above equation as follows 
\begin{align} \label{eq:mixed_partial_ders}
0 &> \frac{\partial^2}{\partial n \;\partial k} \Delta_{\mathrm{CE}}  \; = -n^{-k} \Big( 1 + (1-k)\log n\Big) \overline{Y}, \\
0 &< \frac{\partial^2}{\partial n \;\partial \overline{Y}} \Delta_{\mathrm{CE}} \; = (1-k)n^{-k}. 
\end{align}
Since we are assuming that $\frac{\mathrm{d}k}{\mathrm{d}t}<0$ and $\frac{\mathrm{d}\overline{Y}}{\mathrm{d}t}>0$ hold over the relevant time frame, then the above expressions for the mixed partial derivatives will ensure that both terms on the RHS of Eq. \eqref{eq:mixed_der_inflation_wrt_n} will be positive, and that
\begin{align}
  \frac{\partial}{\partial n} \left( \frac{\mathrm{d} }{\mathrm{d} t}  \Delta_{\mathrm{CE}} \right) &> 0,
\end{align}
will alway hold over the same time period. In other words, increasing the level of diversity, $n$, will increase the rate of change (i.e., the slope) of the CE inflation over time.

\begin{figure}[h]
  \centering
  \includegraphics[width=0.75\textheight]{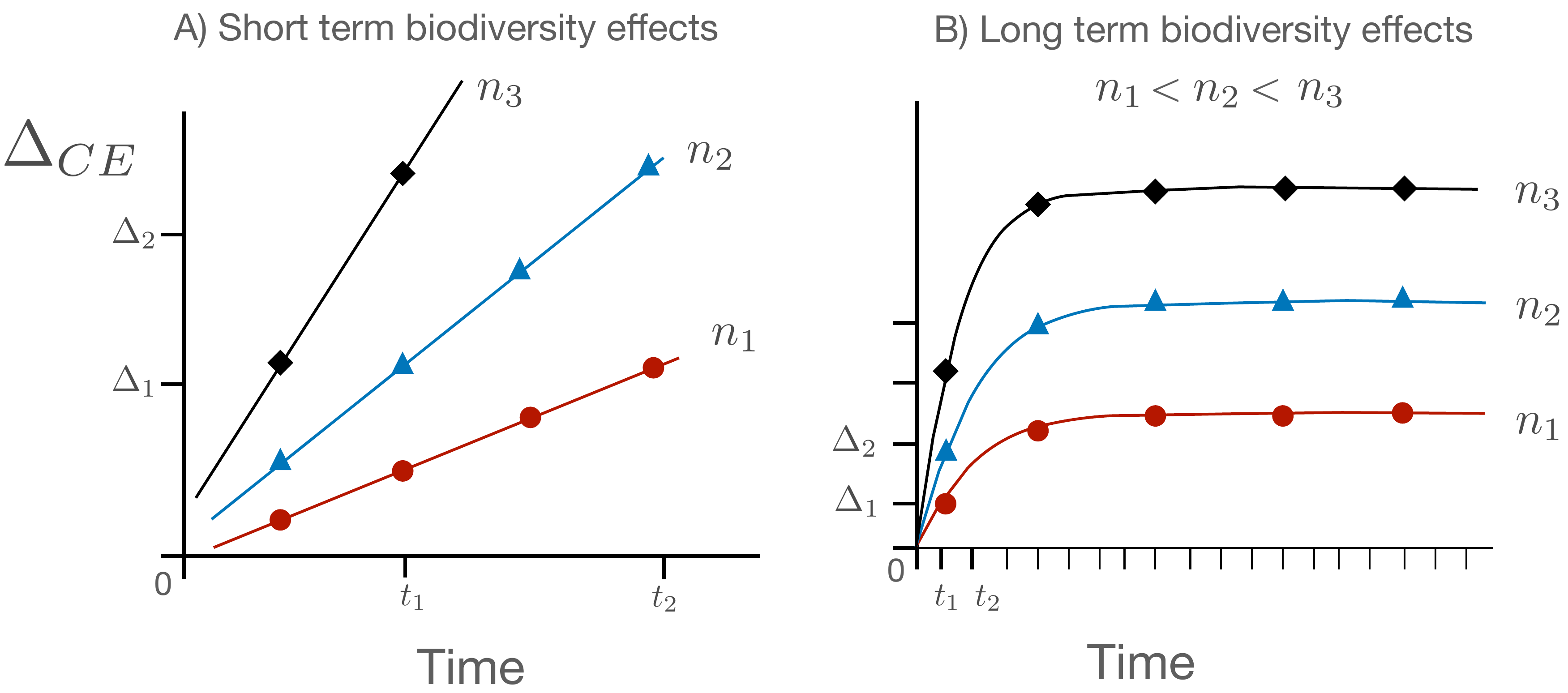}
  \caption{Short and long term biodiversity effects (CE)}{}
  \label{Fig:yields_time}
\end{figure}

Recall from Eq.  \eqref{eq:mixed_der_inflation_wrt_n} that $\frac{\partial}{\partial n}  \Delta_{\mathrm{CE}} > 0$. What this means is that increasing diversity should increase the inflated CE effect at any given moment in time, a result that can be illustrated by $ \Delta_{\mathrm{CE}}$ curves being shifted upwards with increases in diversity, $n$ (Fig. \ref{Fig:yields_time}A and B). From expression \eqref{eq:mixed_der_inflation_wrt_n} above we can also see that the more species in the mixture (i.e., the higher the  diversity $n$), the greater this increase in the inflationary effect is over time, as indicated by the increase in the slope of each curve as $n$ increases  (Fig. \ref{Fig:yields_time}A). Had biodiversity no effect on how the inflation of CE changes over time, then we would expect to see the lines in Fig. \ref{Fig:yields_time}A run parallel to each other rather than see the slopes increase with greater $n$.

CE inflation increases over time, and increasing diversity (greater $n$) amplifies this effect by increasing the slope of the curve describing how $\Delta_{\mathrm{CE}}$ changes across time  (Fig. \ref{Fig:yields_time}). Note that Fig. \ref{Fig:yields_time} is exactly the result observed and reported by \citet{reich2012} -- in particular, their Figure 2A \citep{reich2012}, which shows an identical pattern to our Fig \ref{Fig:yields_time}. However, \citet{reich2012} assumed that this pattern, whereby an increase in the slope of the ecosystem effect across time as diversity increases,  is an effect of how BEF operates over time, while we were able to to demonstrate, based on simple first principles, that this pattern will emerge simply as a \emph{monoculture} effect.  With no diversity effect all lines in Fig \ref{Fig:yields_time}b will be parallel (slopes all equal).

Over long periods when the change in $k$ and $\overline{Y}$ come to a halt or slow down (i.e., outside the relevant time frame where $\frac{\mathrm{d}k}{\mathrm{d}t} < 0$ and $\frac{\mathrm{d}\overline{Y}}{\mathrm{d}t} > 0$), the change in $\frac{\mathrm{d}}{\mathrm{d}t}\Delta_{\mathrm{CE}}$ should slow down and stop; that is $\frac{\mathrm{d}}{\mathrm{d}t}\Delta_{\mathrm{CE}} \rightarrow 0$ as $\frac{\mathrm{d}k}{\mathrm{d}t} \rightarrow 0$ and $\frac{\mathrm{d} \overline{Y}}{\mathrm{d}t} \rightarrow 0$. The effect of diversity on the rate of change of artificial inflation will cease as all lines flatten and become parallel (Fig. \ref{Fig:yields_time}b). (Important note: depending on the property of interest, the degree the monoculture property shifts over time and the the time scales these shifts occur will vary widely.) This longer term effect, predicted here on first principles as a simple monoculture effect  (Fig. \ref{Fig:yields_time}b), yet again reproduces another experimental pattern that was observed and attributed to the effects of biodiversity by \citeauthor{reich2012}  \citep[see Figure 3b in][]{reich2012}.

Although some possibility for controlling this pseudo-biodiversity effect may exist when measuring how the net biodiversity effect, $\Delta_{\mathrm{Net}}$, changes over time, this will not the case for the Complementarity Effect, $\Delta_{\mathrm{CE}}$, given the non-identifiability of different components of the LH partitioning (except in the special case where the artificially inflated selection effect is zero because the exponent parameter, $k$, and the total yield of all species, $Y$, are the same; see \citealt{pillai.gouhier_19}). 

\section{Pseudo-biodiversity effects across space: the role of \emph{independence} and arithmetic effects}
In the previous section we demonstrated  how \emph{population-level} effects can be easily misconstrued as biodiversity effects across temporal scales. We will now consider biodiversity changes across spatial scales \citep[as for example in][]{qiu.cardinale_20} in order to study the role of two other pseudo-biodiversity effects: \emph{species independence} effects and \emph{arithmetic} effects, both of which can similarly be mischaracterized as biodiversity effects.

Starting in a single site, recall that the artificially inflated net biodiversity effect and total biodiversity effect when all species have the same $k$ parameter value is
\begin{align}
  \Delta_{\mathrm{Net}} &= \left( n^{1-k} -1 \right) \overline{Y}  \label{eq:site_delta}\\
  y_{\mathrm{_{Total}}} &= n^{1-k}\overline{Y} \label{eq:site_Y}
\end{align}

In a metacommunity with a total number of $m$  sites, we can calculate the aggregate biodiversity effect as a simple sum of local effects. This aggregate will be defined as the sum of biodiversity effects \emph{across the metacommunity} -- and can be calculated for both the artificially inflated portion of the total effect ($\sum_j \Delta_{ {\mathrm{Net}}_j }$), and the total biodiversity effect ($\sum_j  y_{_{\mathrm{Total}_j } } $) by summing across each  site $j$ in the metacommunity, respectively:
\begin{align}
  \sum_{j\in \mathcal{M}} \Delta_{ {\mathrm{Net}}_j } &= \sum_{j\in \mathcal{M}} \left( n_j^{1-k} -1 \right) \overline{Y}_j,  \label{eq:sum_Y_pre}\\
  \sum_{j\in \mathcal{M}}  y_{_{\mathrm{Total}_j } }  &=  \sum_{j\in \mathcal{M}} n_j^{1-k}\;\overline{Y}_j.  \label{eq:sum_delta_pre}
\end{align}
where $\mathcal{M}$ is the index set for a metacommunity of size or scale $\left| \mathcal{M} \right|= m$; that is, a metacommunity with a total of $m$ sites.

In order to investigate \emph{independence} and \emph{arithmetic} effects  we will first control for ecosystem functioning that arises from species differences or from possible variation among communities as one sums \emph{across} the metacommunity, as would be seen in expressions \eqref{eq:sum_delta_pre} and \eqref{eq:sum_Y_pre} above. To do so we will make the following set of Assumptions:
\begin{tcolorbox}[breakable, enhanced]\label{Key Assumptions}
\centerline{\textbf{Key Assumptions}} 
\begin{assumption} \label{a1}
  All sites have identical richness, $n$.
\end{assumption}
\begin{assumption}\label{a2}
  All species have the same value for the exponent parameter $k$ (an assumption made previously).
\end{assumption}
\begin{assumption}\label{a3}
  All species have the same monoculture yields within sites, $Y$, and this monoculture yield $Y$ does not vay between sites.
\end{assumption}
\begin{assumption}\label{a4}
  All species in mixture are competitively neutral and their abundance  is simply their monoculture abundance multiplied by their relative frequency in mixture $\frac{1}{n}$ (an assumption made previously).
\end{assumption}
\end{tcolorbox}
Assumptions \ref{a1}-\ref{a4} mean that all species are effectively functionally equivalent, and that all sites are essentially identical (for our purposes, at least). That is, with these assumptions we have a homogeneous and uniform biome at the metacommunity scale, with no variation across the region. We will now see how \emph{independence} and Arithmetic effects arise and can be mischaracterized as true or meaningful biodiversity effects across spatial scales.

\subsection{Species Independence effects}
Species \emph{Independence effects} relate to the measurement of the biodiversity effects (ecosystem functioning) of sets of species that exist independently of each other due to, for example, being spatially separated in different habitats. The effect arises when one ignores the inevitable turnover in species (or other OTUs) as one moves to larger spatial scales. 

In order to more clearly isolate the specific role of \emph{independence effects} in biodiversity-ecosystem studies we will add two supplementary Assumptions. First, if we make a supplementary assumption that all species have a linear abundance-functioning relationship so that $k=1$, then we can remove role played by Population-level effects (which we already investigated previously) leaking into our calculation of biodiversity effects explored in the previous section.
\begin{tcolorbox}[breakable, enhanced]
\begin{assumption} \label{a5}
  The population exponent parameter for all species is $k=1$ (linear monoculture curve).
\end{assumption}
\end{tcolorbox}
With the above Assumption \ref{a5}, the artificial net biodiversity effect caused by the leakage of population-level effects becomes zero within each site, and thus, the sum of the inflated net effect across all sites vanishes:
\begin{align} \label{eq:sum_delta_zero}
  \sum_{j\in \mathcal{M}} \Delta_{ {\mathrm{Net}}_j } &= m\left( n^{1-k} -1 \right) \overline{Y} = 0.
\end{align}
This means that under our assumptions there will be no measurable net biodiversity effect within sites regardless of how many species are added. Combining Assumption \ref{a5}  with the previous four Assumptions, we now not only have a regional biosphere that is homogeneous/uniform across space, but also a metacommunity system where species addition or subtraction will have \emph{no} effect on ecosystem functioning, either locally regionally. 

In the case of the \emph{total} biodiversity effect (and not just that portion of the biodiversity effect inflated relative to the null), when we sum across all local sites (taking into consideration Assumptions \ref{a1}-\ref{a4}) we get
\begin{align} \label{eq:sum_Y}
  \sum_{j\in \mathcal{M}}  y_{_{\mathrm{Total}_j } }  &= m\;n^{1-k}\;\overline{Y} =  m\overline{Y}. 
\end{align}

Both Eqs. \eqref{eq:sum_delta_zero} and \eqref{eq:sum_Y} show the biodiversity effects (both the total and the artificial net effect) that represent trivial effects arising from simply summing the effects at local sites across a metacommunity. As such, these expressions provide a baseline for investigating effects that are not simply a consequence of summing local effects.

Also note how both the expressions for Eqs. \eqref{eq:sum_delta_zero} and \eqref{eq:sum_Y} do not depend on how we arbitrarily classify or group functionally equivalent species within sites. Both Eqs. \eqref{eq:sum_delta_zero} and \eqref{eq:sum_Y}  hold even if Assumption \ref{a1} is violated -- that is, if diversity varies between sites ($n_j$ differences between each site) -- just so long as the other Assumptions hold. At the same time, these assumptions ensure that the same species can appear in different sites without affecting results obtained from Eqs. \eqref{eq:sum_delta_zero} and \eqref{eq:sum_Y}. In other words, the type of species turnover, from complete turnover (all sites with different species), to no turnover (all sites with the same species), will not affect the sum of the the net and total biodiversity effects across sites.

We now turn to our explicit definition of Independence effects:
\begin{definition}[Independence Effects]\label{d2}
  Effect arising solely due to species habitat occupancy being (spatially) segregated. That is, due to non-interacting species being incapable of occupying the same spatial habitat, resulting in turnover (e.g., lions and polar bears).
\end{definition}

Now that we have a definition of \emph{independence effects}, we will make use of a second Supplementary Assumption that will ensure that we have maximum species turnover between sites:
\begin{tcolorbox}[breakable, enhanced]
\begin{assumption} \label{a6}
  Each species can only occupy one site in the metacommunity. That is, there is complete turnover between sites with no shared species between them.
\end{assumption}
\end{tcolorbox}
The purpose of Assumption \ref{a6} is to provide an idealized scenario which will allow \emph{independence effects} arising from perfect species turnover to be isolated and measured without being obscured by the confounding effects that arise when sites have overlapping species. This will become more clear next when, in contrast to summing up biodiversity effects in individual sites (effects measured \emph{across the metacommunity}), we measure biodiversity effects \emph{at the metacommunity or regional scale}.

Recall above how we referred to the biodiversity effects that arise \emph{across} the metacommunity as that which arises when local effects are trivially summed up across sites. Now we will consider biodiversity effect that occur \emph{at} the metacommunity scale -- that is, biodiversity effects that are measured when the larger metacommunity scale is treated as a single site.  Using the full range of Assumptions \ref{a1}-\ref{a6} we will now see how measuring the effects of the total and the net biodiversity effects \emph{at} the metacommunity scale can give rise to \emph{independence effects}.

\subsubsection*{Artificial net biodiversity effect arising from independence }
The LH Net biodiversity effect \emph{at} the metacommunity scale is simply $\text{Total observed effect} - \text{LH null effect}$. To measure the the Net effect \emph{at} the metacommunity scale we first need to determine what the LH null or expected baseline is at the metacommunity scale. Given that the metacommunity species richness is, according to our assumptions,  $m\times n$, the LH null expectation is simply
\begin{align} \label{eq:metacomm_null}
  \text{Expected (LH baseline)} &= \frac{1}{mn} \sum_{j\in \mathcal{M}} \sum_{i=1}^n Y_{i,j}  \notag \\
  &= Y. 
\end{align}
Also, given our assumptions (in particular, \ref{a5}), our total observed biodiversity effect \emph{at} the metacommunity scale, $y_{_\mathcal{M}}$, will simply be the quantity in Eq. \eqref{eq:sum_Y}:
\begin{align} \label{eq:y_at_metacomm}
  y_{_\mathcal{M}} &= \sum_{j}  y_{_{\mathrm{Total}_j } } = m\;Y 
\end{align}

Under our simple model assumptions, it is apparent that the entire net biodiversity effect that will be measured \emph{at} the metacommunity scale will be solely due to species \emph{independence effects}, and that this measured effect will thus represent yet another type of artificial inflation of the Net biodiversity effect, this time one occurring at the metacommunity spatial scale $\Delta^{^\mathcal{M}}_{\mathrm{Net}}$:
\begin{align} \label{eq:net_at_metacomm}
  \Delta^{^\mathcal{M}}_{\mathrm{Net}} &=\text{Observed} - \text{LH null}  \notag \\
  &= \left(m\;n^{1-k}-1\right) Y.  \notag \\
  &= \left(m -1\right) Y \;\;\;\;\;\;\text{(given $k=1$) }. 
\end{align}

When studying \emph{independence effects} we will also try to control for the extensive nature of some types of ecosystem properties (e.g., biomass/yield), whose observed increases at larger spatial scales may simply be the result of them being additive functions of the number of sites sampled. We can do this here by dividing $\sum_{j}  y_{_{\mathrm{Total}_j } }$ and $\Delta^{^\mathcal{M}}_{\mathrm{Net}}$ by $m$ to get
\begin{align}
  & \frac{1}{m}\sum_{j}  y_{_{\mathrm{Total}_j } }  = Y,  \label{eq:Y_scaled} \\
  & \frac{1}{m}\Delta^{^\mathcal{M}}_{\mathrm{Net}} =  \left(1 - \frac{1}{m}\right) Y.  \label{eq:Delta_scaled}
\end{align}
Controlling for $m$ allow us to eliminate the most intuitive and self-evident spatial effect of all: cases where increasing sample space increases the thing measured in an additive manner. In other words, it allows us to control for the trivial increase in biodiversity effects that occurs at larger spatial when a property is an \emph{extensive} quantity.

\begin{figure}[h]
  \centering
  \includegraphics[width=0.70\textheight]{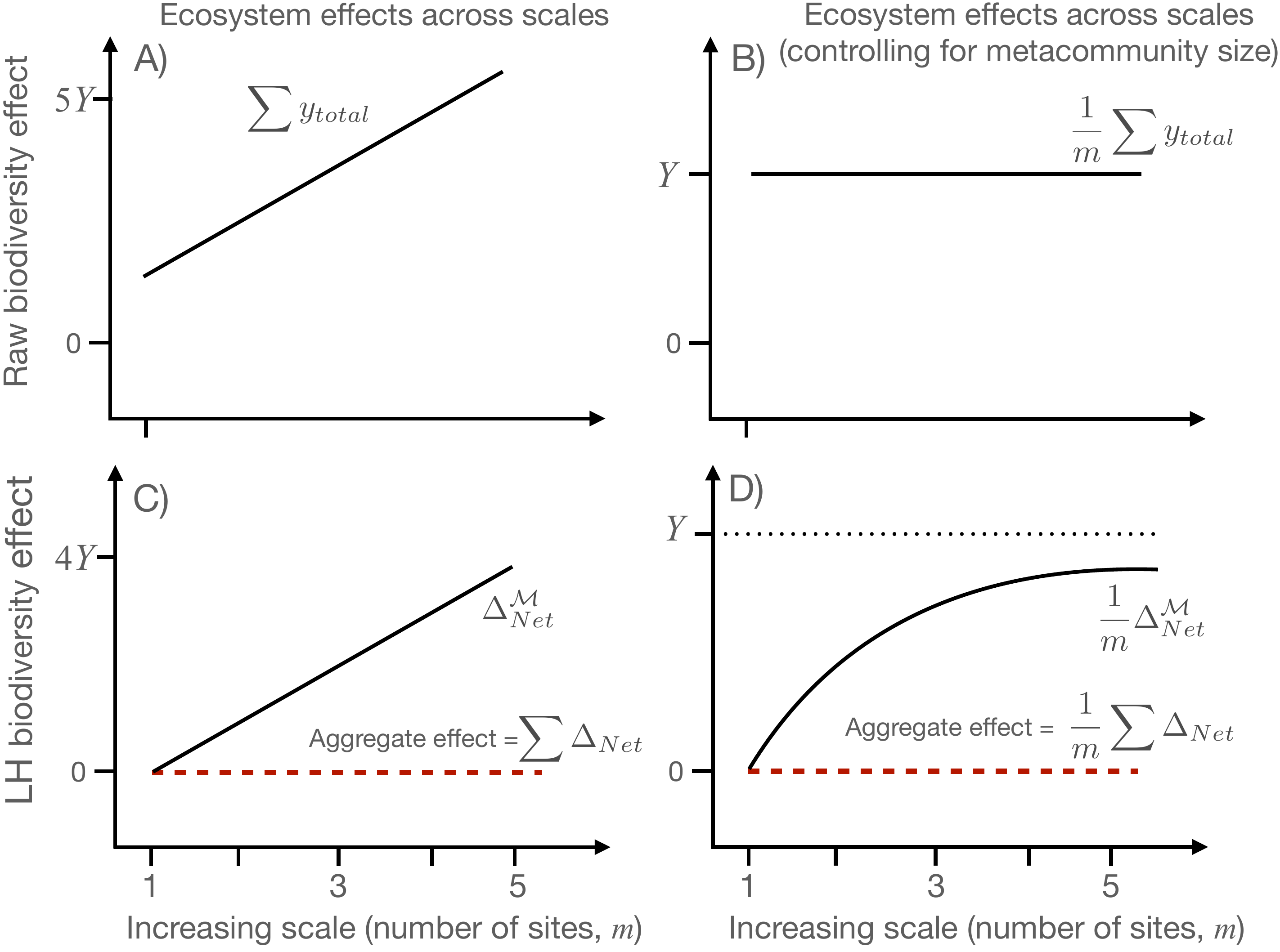}
  \caption{Independence effects}{}
  \label{Fig:independence_effects}
\end{figure}

Note that the \emph{total} biodiversity effect \emph{at} the metacommunity scale (Eq. \eqref{eq:y_at_metacomm}) is equivalent to the \emph{total} effect \emph{across} the metacommunity, $y_{_\mathcal{M}} = \sum_{j}  y_{_{\mathrm{Total}_j } } $.  Thus, we would expect a straightforward monotonic increase in total biodiversity effect as spatial scale increases (Fig. \ref{Fig:independence_effects}A). Intuitively (given our set of assumptions), there should be no particular role for the independence effects when measuring the \emph{total} biodiversity effects when summed across sites at the metacommunity scale: total effects will always ultimately reduce to the sum of effects in individual sites, and thus will trivially increase with scale in a linear manner, as expected \emph{regardless} of habitat turnover (Fig. \ref{Fig:independence_effects}A).  By scaling the total biodiversity effect (i.e., controlling for metacommunity size by dividing by $m$), we see that  total yield measured across space does not change, indicating that the independence effect (effect of habitat turnover) had no role in the increase in total biodiversity effect observed at larger spatial scales (Fig \ref{Fig:independence_effects}B).

Similarly, the net biodiversity effect grows in value as spatial scale increases (Fig \ref{Fig:independence_effects}C). Unlike the total biodiversity effect, however, scaling the net biodiversity effect to control for regional size, $\frac{1}{m}\Delta^{^\mathcal{M}}_{\mathrm{Net}}$,  still results in a monotonic increase with $m$ (Fig. \ref{Fig:independence_effects}D). This indicates that the observed increase in net biodiversity effect in Fig \ref{Fig:independence_effects}C cannot be reduced to the additive summing of local effects at the regional scale. However, given the assumptions made in our model, the observed increase in the net biodiversity effect can still be attributed to another type of triviality -- specifically, \emph{independence effects} arising from species turnover as one increases spatial scale. The net biodiversity effect is particularly prone to this form of triviality because of the artificially low baseline established by the Loreau-Hector null, which by its very nature cannot incorporate the  effect of species independence and non-interactions that become  \emph{inevitable} as one moves across spatial scales. 

\begin{figure}[h]
  \centering
  \includegraphics[width=0.70\textheight]{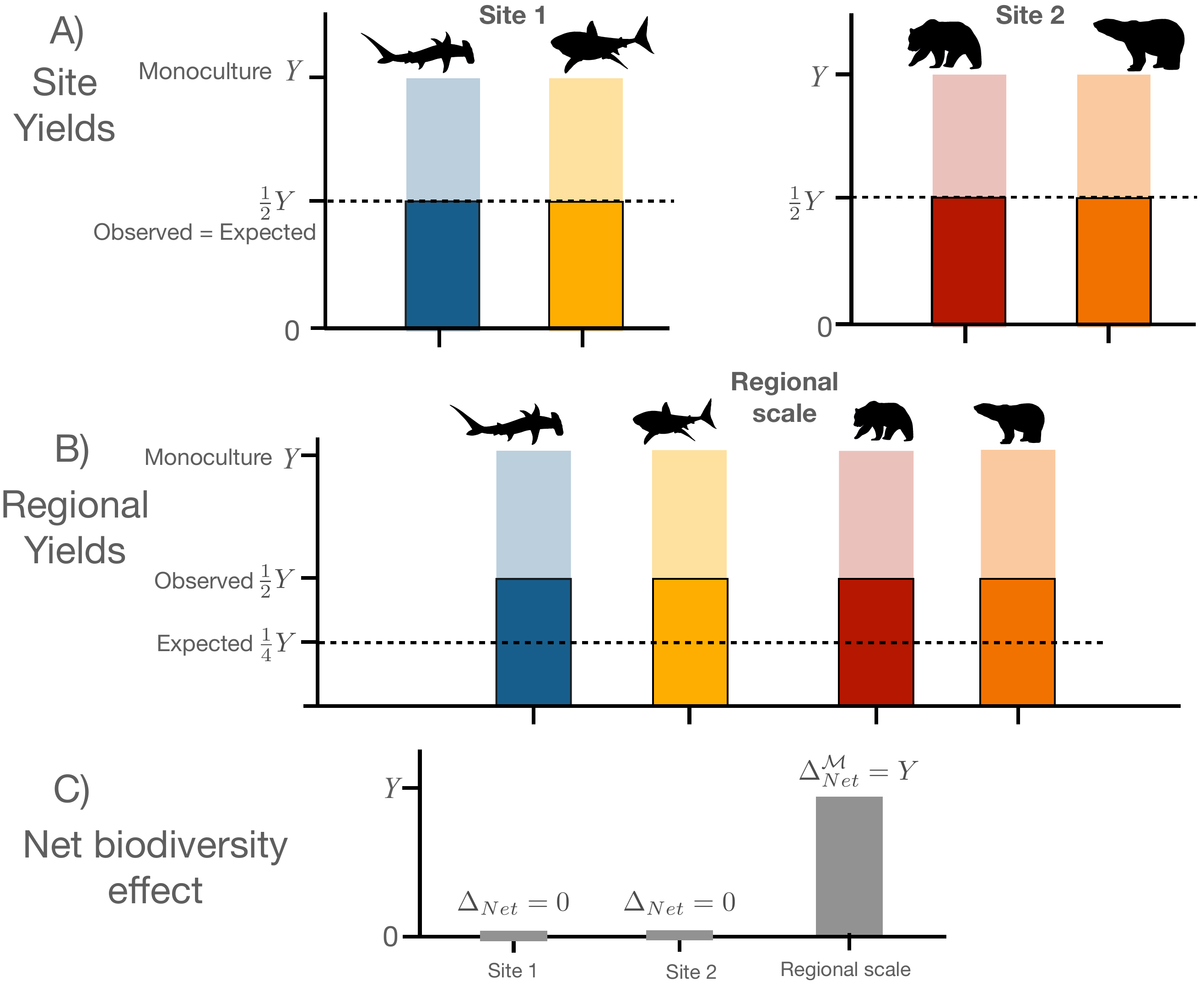}
  \caption{The loreau-hector effect measured at large spatial scales}{}
  \label{Fig:LH_example}
\end{figure}

As an illustration, take a simple (and extreme) scenario involving two spatially separate and distinct habitats (say, a marine and boreal terrestrial habitat), each capable of being occupied by a unique pair of species (Fig. \ref{Fig:LH_example}). In each site the  net biodiversity effect measured by the Loreau-Hector method will quantify the degree to which the corresponding mixture departs from both an expectation of neutrality, and the expectation of there being no positive or negative interaction effects. Keeping to our stated assumptions, each individual site should have a zero net biodiversity effect, and the sum of local effects across the region will also be zero. Each of the four species at the metacommunity scale are obviously incapable of growing across both habitats; yet if we measure the net biodiversity effect at the \emph{regional scale}  a clear net biodiversity effect will be quantified. But this effect is simply a consequence of an absurd assumption: that all species when grown separately  (in monocultures) are \emph{by default},  assumed to be capable of growing everywhere across the entire region. More so, the LH null is set up in such a way that the monoculture yields that were actually measured were assumed to have applied at a global scale.

The regional scale measure of the net biodiversity effect illustrated here is not `incorrect', it is just trivial. It is simply an artefact of the fact that \emph{all} species taken on their own, cannot potentially exist \emph{everywhere} in the world. To some degree the LH method, when applied at ever increasing spatial scales, is merely quantifying the result of this simple truism. As scale increases across gradients and even biomes, the greater the chance that the biodiversity effects observed are the predominantly trivially true effects arising from simple independence effects. Although the \emph{total} biodiversity effect may not be applied across scales in the manner shown above, the Loreau-Hector net effect has already begun to be applied in a way without accounting for these trivial effects \citep[e.g.,][]{qiu.cardinale_20}.

Note that under our idealized assumptions the artificial inflation of the net biodiversity effect due to independence  in Eqs. \eqref{eq:net_at_metacomm} and \eqref{eq:Delta_scaled} are not affected by the number of species in each site. That is, even if each site contains only one unique species ($n=1$), then it becomes unambiguously clear/obvious how the net biodiversity effect measured will solely be due to species only being capable of occupying separate spaces without interacting. The cases considered here are extreme scenarios that allow us to isolate and illustrate this trivial effect, one that would otherwise be missed when confounded with real biodiversity effects in natural experiments.

Unless one assumes axiomatically that every species can on their own exist everywhere on the globe, this quantified `phenomena' is obvious,  and in-of-itself not a meaningful biological result. If on the other hand one assumes \emph{axiomatically} that not all species can potentially exist everywhere on the globe, then this independence effect will eventually arise an inevitable confounding factor. All things being equal -- that is, even in a world where all species are functionally equivalent within their habitats, and all habitats and regions are functionally identical to each other with no interactions between them, such that some biologically induced property is invariant and uniform across space  -- one will still always measure a net biodiversity effect across the homogenous landscape merely by increasing the spatial scales of observation.

\subsubsection{Independence effects arising from taxonomic resolution}
A version of the independence effect arises when we attempt to quantify effects at different levels of taxonomic resolution. We assumed throughout this paper that all species are essentially indistinguishable. What this means is that the artificial net biodiversity effect arising from independence effects is also in part an artefact of how species are arbitrarily classified, or the degree of taxonomic resolution. If we decide to artificially and arbitrarily resolve a homogenous biological entity across space into smaller operational taxonomic units (OTUs), we can artificially create biodiversity driven ecosystem effect  that is simply an artefact of how we decide to divide, classify and order the biological world. 

\begin{figure}[h]
  \centering
  \includegraphics[width=0.70\textheight]{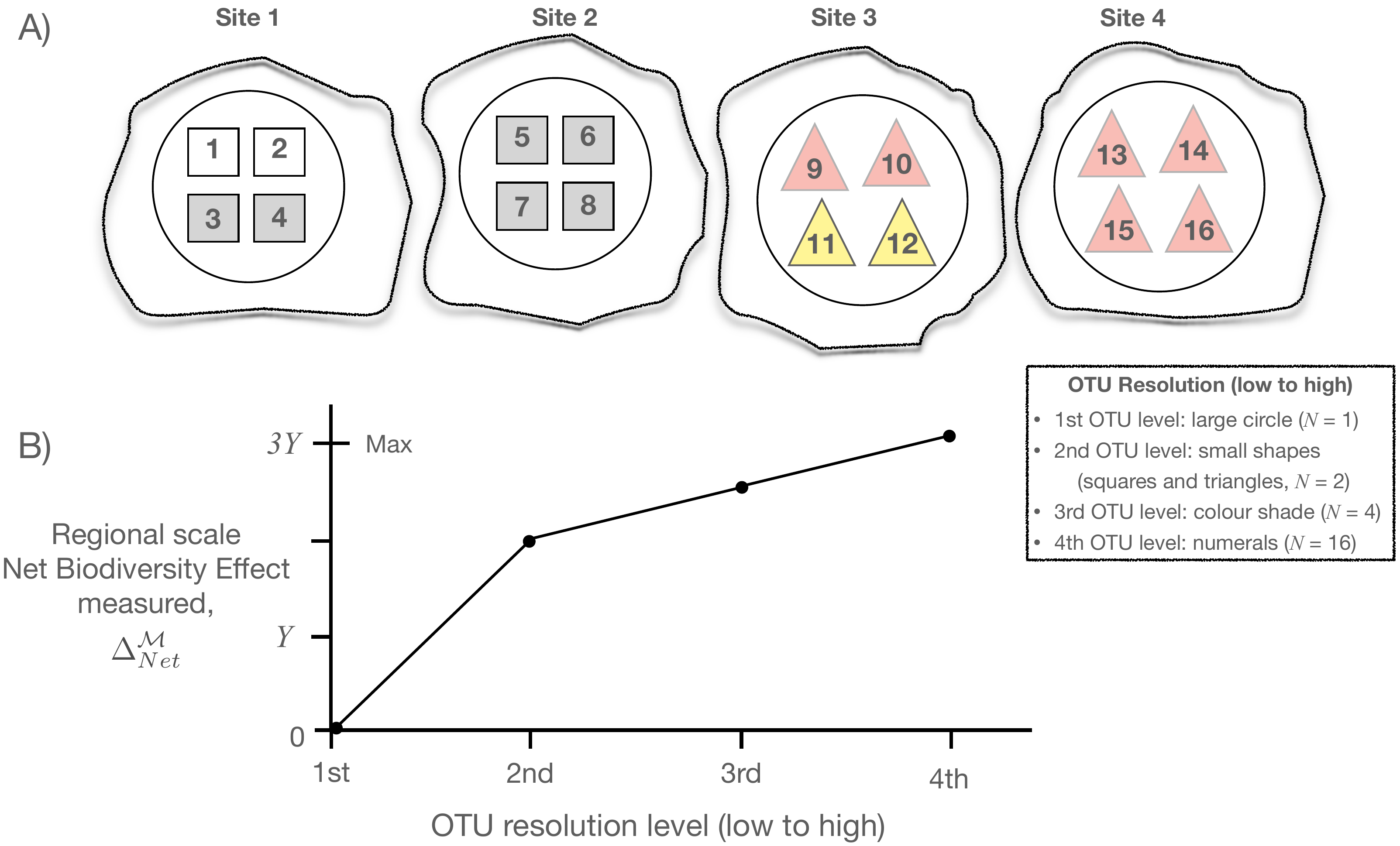}
  \caption{Independence effects arising from arbitrary classification of OTUs}{}
  \label{Fig:OTU_resolution}
\end{figure}

We can illustrate how different degrees of OTU resolution (Fig. \ref{Fig:OTU_resolution}A) will give rise to different measurements of ecosystem functions based solely on how one wished to divide or classify entities (Fig. \ref{Fig:OTU_resolution}B). If we assume again that all OTUs within a given site have the same monoculture yields, $Y$, then we can plot the net biodiversity effect $\Delta^\mathcal{M}_{\mathrm{Net}}$ using a simple formula: $\Delta^{^\mathcal{M}}_{\mathrm{Net}}= \left( m-\overline{\nu}\right)Y$. Here $\overline{\nu}$ is the average number of sites that a species/OTU occupies in the metacommunity  (note how we are temporarily relaxing for our purposes here the assumption that each species/OTU only occurs in a single site; see Appendix \ref{otu_resolution} for details).

At the coarsest level of resolution (represented by the large circles in \ref{Fig:OTU_resolution}A) there is only one biological entity across all sites (regional number of OTUs $N=1$), and the net biodiversity effect \emph{at} the regional scale is zero ($\Delta^\mathcal{M}_{\mathrm{Net}}=0$, as one would expect). If we resolve this entity into the smaller OTUs represented by simple shapes (squares in Sites 1 and 2, and triangles in Sites 3 and 4), then the number of regional OTUs is $N=2$ and the net biodiversity effect measured at the metacommmunity scale is $\Delta^\mathcal{M}_{\mathrm{Net}}=0$, which can be calculated directly by inspection, or by using the formula we described above (see Appendix \ref{otu_resolution}). Similarly, arbitrarily increasing the OTU resolution (e.g., colour or numbers) will potentially give rise to different net biodiversity effect measurements. 

As we see in this example, changing OTU classification gives rise to an independence effect because of how OTU reclassification/resolution affects the degree of turnover.  Notice how if classification leads all OTUs to appearing everywhere, then the net biodiversity effect at the metacommunity scale vanishes:  $\Delta^{^\mathcal{M}}_{\mathrm{Net}}= \left( m-\overline{\nu}\right)Y=0$, since average site occupancy $\overline{\nu}$ becomes equal to the number of sites $m$.

\subsection{Arithmetic effects}
Another trivial effect that can potentially be mischaracterized as a meaningful biodiversity effect are those arising simply from application of  arithmetic or mathematical operations across different scales. We already saw an example of this with effects arising from simply summing some property \emph{across} metacommunities (``the more things I count, the more things there are''). This gives us the following simple and suitably general definition

\begin{definition}[Arithmetic Effects]\label{d3}
  Effect that are simply artefacts of applying mathematical operations at or across different scales. That is, observations or measurements that simply arise when functions are applied at different spatial domains, but that are not due to any biological feature of the system.
\end{definition}
As mentioned, a self-evident and trivial example of arithmetic effects that we discounted immediately was the increase in biomass one observes naturally as one increases the number of sites over which one is measuring across. We controlled for this previously by dividing both the total and net biodiversity effects on yield by the number of sites, $m$ (Eqs. \eqref{eq:Y_scaled} -\eqref{eq:Delta_scaled}). A slightly more subtle category of spurious quantification arises with the non additivity of some functions being applied across the variables in a domain ,
\begin{align} \label{eq:non_additive}
  f\left(x_1+x_2+ \cdots + x_n\right) & \neq f\left(x_1 \right) + f\left(x_2 \right) + \cdots + f\left(x_n \right). 
\end{align}
Applying operations or functions to elements in a domain that are aggregated differently will give different results. Thus the differences measured at different aggregations might have no biological meaning whatsoever.

A good illustration of the quantification of spurious arithmetic effects based on the non additivity of some function or parameter across spatial scales is given by a recent study exploring how the biodiversity-functioning relationship changes with scale (Thompson et al. 2018). Here an attempt was made to see how the measurement of the exponent in a biodiversity-ecosystem saturation curve will change as one increases the number of sites over which it is measured. (Note: a saturating curve will require here relaxing Assumption \ref{a4}, stating that all species are competitively neutral and overlap perfectly in habitat niche use, and Assumption \ref{a5}, which states that the exponent by which yields are scaled in local mixtures will always be $\beta = (1-k) =0$.)

In a single site $j$,  where the average species monoculture property (e.g., yield) is given by coefficient $Y_j$,  the total yield will scale with diversity, $n_j$, by $Y_j \times n_j^{\beta_j}$ (for exponent $\beta_j$). Summing across all sites in a metacommunity defined by the set $\mathcal{M}$ gives the metacommunity or regional scale version of the saturation curve,
\begin{align} \label{eq:saturation_metacomm}
   Y_{_\mathcal{M} } \; N^{{ \beta_{\mathcal{M}}} }  &= \sum_{j\in\mathcal{M}} Y_j \; n_j^{\beta_j}, 
\end{align}
where the metacommunity diversity $N$ is the sum of all local diversity levels, $N =\sum_j n_j$. On the LHS of the above expression  $Y_{_\mathcal{M} }$ is the average metacommunity-scale yield of all species, and $\beta_{_\mathcal{M} }$ is the scaling exponent at the metacommunity-scale. 

For the purpose of this analysis we will assume that for all the species in each of the $m$ sites that both the  monoculture coefficients of each species and the saturation exponent in each site will be the same, and will be given by $Y_j=Y$ and $\beta_j=\beta$, respectively. Furthermore, it will be assumed that there is complete turnover between sites (no species overlap), and that each site has the same diversity level, $n$. The metacommunity parameters $Y_{_\mathcal{M} }$ and $N$ then reduce to $Y_{\mathcal{M} } = Y$, and $N=m\times n$ (See Appendix A for discussion of general case).

The purpose of the original \citet{thompson.isbell.ea_18} study was to attempt to see (via simulations) whether the saturation exponent at the metacommunity scale differed from the saturation exponent within local sites, and if so, how it differed.  However, simple rearrangement of Eq. \eqref{eq:saturation_metacomm} will give the explicit value of the metacommunity saturation exponent in terms of the local exponent,
\begin{align} \label{eq:beta_M}
   \beta_{_\mathcal{M} } &= \frac{\beta \log n \;+ \log m \; }{ \;\;\;\log n \;+ \log m \;}  \;\;\;\;\;\; (\text{for} \;\;\; m\times n \neq 1). 
\end{align}
The real question that arises with this type of demonstration then, is why it was originally thought biologically meaningful (or even appropriate) to investigate whether an exponent measured at the aggregate of multiple sites differed from that which was measured within individual sites. By the elementary rules of arithmetic, the exponents, in general, had to be different (we put aside here the fact that the authors appear not to have been aware that one could easily write out the explicit formula for exponent $\beta_{_\mathcal{M} }$, let alone the inevitability of the arithmetic relationship of non identity; see Appendix \ref{saturationexponent}).

That the exponent found at regional scales is different, in general, from exponents found at local scales is  not in-of-itself a ``biodiversity effect'' at the regional scale, but a simple arithmetic effect arising from the difference between the sum of functions (exponentials) and the function (or exponentiation) of a sum. By the rules of simple arithmetic (in this case, the rules determining the order of operations) it \emph{has} to be different (unless, of course, the function is linear). The changes in a measured property as one moves from individual entities to aggregates is not in-of-itself a biological, let alone an ecological feature of a shift in scales; in cases like this it will always apply, whether for penguins or paperclips.

This change in parameter across scales is simply an arithmetic property of the nonlinearity or non-additivity of certain types of mathematical operation. More importantly, this exponent parameter does not tell us anything meaningful at large spatial scales: it does not indicate the strength of the biodiversity-ecosystem functioning relationship -- more specifically, it does not tell us how ecosystem functioning changes or saturates at large spatial scales at all. To see this let us return to the metacommunity saturation expression in Eq. \eqref{eq:saturation_metacomm}. Recall that $\beta_{\mathcal{M}}$ is the exponent measured at the regional scale (that is, the exponent estimated as though the regional scale were a single site). Let us keep to all the assumptions made earlier (complete species turnover, same diversity levels in all sites, etc.).

We now have a total biodiversity saturation effect measured \emph{at} the metacommunity scale, as given by the LHS of expression Eq. \eqref{eq:saturation_metacomm}:
\begin{align}\label{eq:saturation_at_m}
  y_{_{\mathcal{M}} } &= Y_{_\mathcal{M} } \times N^{{ \beta_{\mathcal{M}}} }. 
\end{align}
We also have a total raw biodiversity saturation effect obtained by summing across within-site yield totals across the $j$ sites in the metacommunity, as given by the RHS of Eq. \eqref{eq:saturation_metacomm}:
\begin{align}\label{eq:saturation_across_m}
  \sum_j y_{_{\mathrm{Total},j}} &= m \times Y \times n^{\beta}.
\end{align}
Clearly both Eqs. \eqref{eq:saturation_at_m} and \eqref{eq:saturation_across_m} above are equal to each other, by definition.

An experimenter studying our system without being aware of the underlying structure of the metacommunity (e.g., species turnover rate, etc.) can still estimate the saturation exponent $\beta_{\mathcal{M}}$ at the regional scale by using Eqs. \eqref{eq:saturation_at_m},
\begin{align}\label{eq:beta_measured}
  \beta_{\mathcal{M}}^* &=  \frac{  \log \left( \frac{1}{Y_{_\mathcal{M} }} \;  y_{_\mathcal{M}} \right) }{\log N} =  \frac{  \log \left( \frac{1}{Y_{_\mathcal{M} }} \sum_j y_{_{\mathrm{Total},j}} \right) }{\log N}. 
\end{align}
Here we use an asterisk, $*$, in the parameter symbol $\beta_{\mathcal{M}}^*$ to indicate that this an \emph{observed} or experimental estimate of $\beta_{\mathcal{M}}$. This estimate was made using the total yields observed across an entire metacommunity that was treated as a single site. As such, the exponent $\beta_{\mathcal{M}}^*$ is being treated by our purported experimenter/observer as a fixed, \emph{non-emergent} parameter observed at this regional `site'. In other words, $\beta_{\mathcal{M}}^*$ is not being considered as though it were a function of some more basic underlying metacommunity parameters.

By knowing the total biomass/functioning and the species richness at the regional scale it appears that one can estimate a key parameter, $\beta_{\mathcal{M}}^*$, determining the form of the biodiversity-ecosystem relation. However, the biological nature of this purported saturation exponent is only apparent. Estimates of $\beta_{\mathcal{M}}^*$  will not tell one how ecosystem properties, like yield, changes or saturates with biodiversity at the regional scale in the same way that the $\beta$ exponent does in local sites.

To understand why, first consider how the saturation exponent allows one to see how yield (or some other property) will change as species are added. The \emph{true} effect on yield arising from increases in richness at the regional scale can be found by taking the derivative of $\sum_j y_{_{\mathrm{Total},j}}$ with respect to metacommunity richness $N$. We do this by first using a continuously differentiable version of $\sum_j y_{_{\mathrm{Total},j}}$, and then by noting that $N = m\times n$.  For a given regional scale ($m$) this gives us the \emph{true} increase in yield due to increasing regional diversity (which arises from increasing local diversity),
\begin{align}\label{eq:der_y_sum}
  &\textrm{TRUE regional yield increase given $\beta$}:  \notag \\
  &\frac{\mathrm{d}}{\mathrm{d}N} \sum_j y_{_{\mathrm{Total},j}} =  \beta \times Y \times n^{\beta-1}. 
\end{align}
Notice how this \emph{true} increase in yield based on local saturation curves remains the same across all regional scales (that is, for all values of $m$), given that the expression is independent of the metacommunity size (see Section \ref{true_regional_yield} in Appendix \ref{saturationecurve}). 

However, predicting the effect that regional biodiversity changes will have on regional yields when using the  $\beta_{\mathcal{M}}^*$ obtained from observations at the metacommunity scale, $y_{_\mathrm{Total}}^{\mathcal{M}} = Y_{_\mathcal{M} } \times N^{{ \beta^*_{\mathcal{M}} } }$, will potentially give a very different value. Similar to the TRUE yield increase we can define the EXPECTED increase based on the observed parameter $\beta_{\mathcal{M}}^*$ given a  local exponent $\beta$ (i.e., $\left(\beta^*_{\mathcal{M}} \; | \; \beta \right)$),
\begin{align}\label{eq:der_y_metacomm}
  &\textrm{EXPECTED yield increase based on $\beta_{\mathcal{M}}^*$}:  \notag \\
  & \Bigg[ \frac{\mathrm{d}}{\mathrm{d}N} \; y_{_{\mathcal{M}}} \Bigg]_{ \left(\beta^*_{\mathcal{M}} \; | \; \beta \right) } = {{ \beta^*_{\mathcal{M}} } } \; Y_{_\mathcal{M} } \times (m\;n)^{({ \beta^*_{\mathcal{M}}}  - 1) } . 
\end{align}

For simple saturation curves ($\beta <1$),  it quickly becomes apparent that the EXPECTED change in yield predicted using the observed $\beta^*$ will always be greater than the TRUE change in yield  (except in the special case where the local exponent $\beta$ is equal to 1, or where there is only one site in the metacommunity). That is, the change in yield predicted when using the observed regional exponent $\beta^*_{\mathcal{M}}$  given the local exponent $\beta$ will lead to the relationship
\begin{align}\label{eq:der_y_inequality}
  \textrm{TRUE yield increase} &\leq\; \textrm{EXPECTED yield increase}  \notag \\
  \left[ \frac{\mathrm{d}}{\mathrm{d}N} \sum_j y_{_{\mathrm{Total},j}}  \right]_{\beta} &\leq \; \Bigg[ \frac{\mathrm{d}}{\mathrm{d}N} \; y_{_{\mathcal{M}}} \Bigg]_{ \left(\beta^*_{\mathcal{M}} \; | \; \beta \right) }.  
\end{align}

\begin{figure}[h]
  \centering
  \includegraphics[width=0.70\textheight]{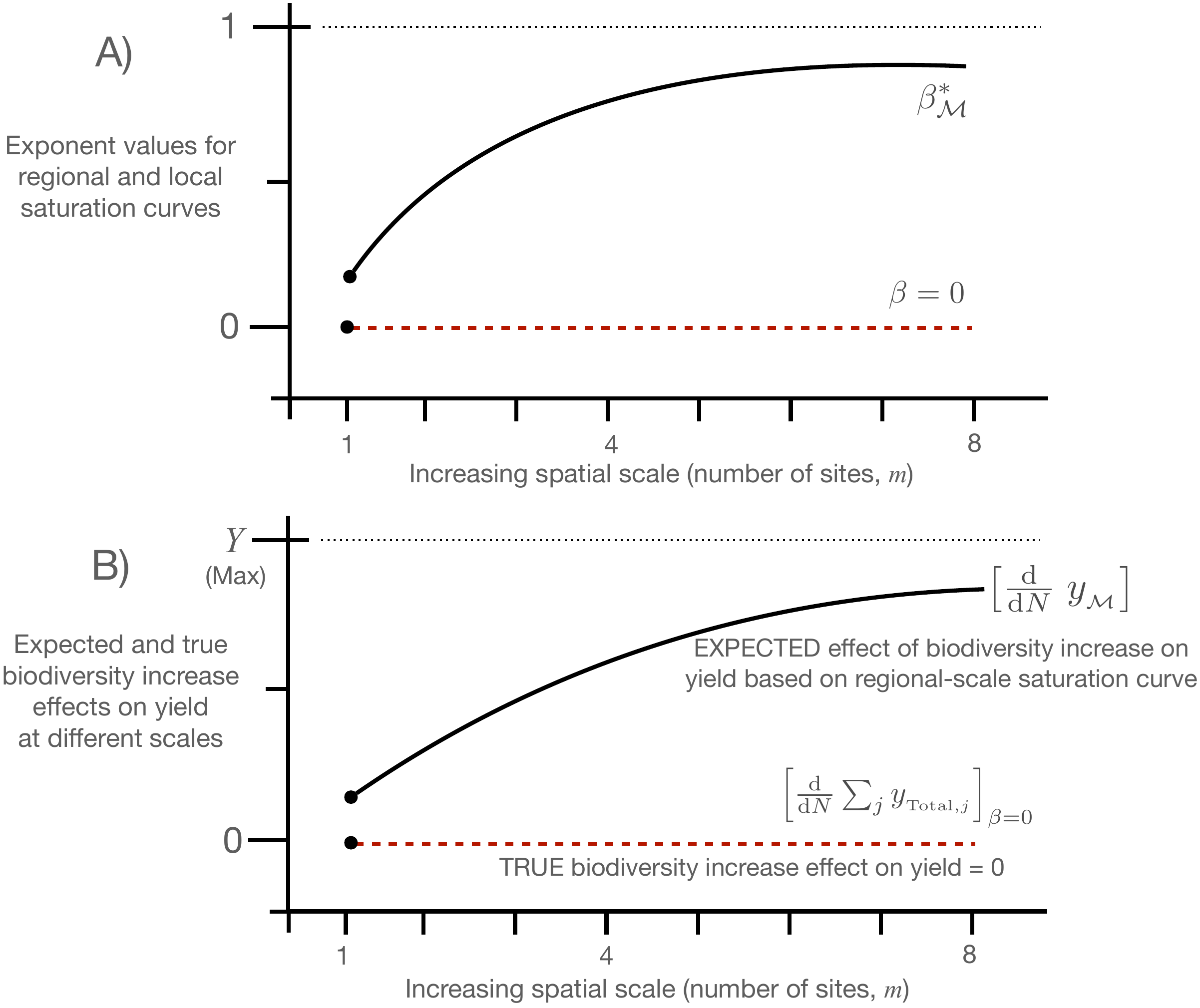}
  \caption{Expected and true effects of species additions on yields based on saturation curves.}{}
  \label{Fig:regional_saturation_curves}
\end{figure}

To better understand the artefactual, non biological character of the regional exponent, consider a special case where the local exponent is set to zero, $\beta =0$. This means that the addition of species to the system will result in there being no local or regional changes in total yield. This corresponds to the special case where all species are neutral within sites, and equally share their habitat niche in a zero-sum manner, $\frac{\mathrm{d}}{\mathrm{d}N} \sum_j y_{_{\mathrm{Total},j}} = 0$. What this means is that no matter to what degree diversity is increased within the metacommunity the total yield will stay constant, \emph{both} at the local and at the regional scale. That is, the true increase yield due to increasing regional diversity, $N$, will be zero:
\begin{align}
  \left[ \frac{\mathrm{d}}{\mathrm{d}N} \sum_j y_{_{\mathrm{Total},j}}  \right]_{\beta=0} &= 0 
\end{align}
Yet, using Eq. \eqref{eq:der_y_metacomm} to predict how regional yield will change with biodiversity using the $\beta^*_{\mathcal{M}}$ parameter observed at the metacommunity scale will give a different, and completely misleading sense of how yield saturates with diversity. In this case we end up with the relation
\begin{align}\label{eq:der_y_greaterthan0}
  0 &\leq \; \Bigg[ \frac{\mathrm{d}}{\mathrm{d}N} \; y_{_{\mathcal{M}} } \Bigg]_{ \left(\beta^*_{\mathcal{M}} \; | \; \beta=0 \right) } 
\end{align}
Despite the \emph{true} change in regional yield, $\frac{\mathrm{d}}{\mathrm{d}N} \sum_j y_{_{\mathrm{Total},j}}$, being zero (because $\beta =0$), the \emph{predicted} change, $\frac{\mathrm{d}}{\mathrm{d}N} \; y_{_\mathrm{Total}}^{\mathcal{M}}$, based on parameters observed at the metacommunity scale (i.e., $\beta^*_{\mathcal{M}}$), will be positive. The end result is a completely erroneous prediction of how yield will change with the addition of species (Fig. \ref{Fig:regional_saturation_curves}). Recall that a saturation curve and exponent parameter are supposed to tell us how yield changes as we increase diversity. Yet, as we have demonstrated here, the saturation exponent obtained at the regional scale is useless for predicting how biodiversity effects saturate regionally -- its very purpose as parameter.

The entire point of the \citet{thompson.isbell.ea_18} paper was to demonstrate that regional saturating exponents would be different from local exponents -- a non scientific question (and one which they answered incorrectly; see also Appendix \ref{saturationexponent}). That the regional exponent parameters are different from local exponent parameters is \emph{not} a biodiversity effect but a simple and obvious arithmetic consequence of the sum of exponentials  not being equal to the quantity obtained when the same bases are first summed, then raised to the exponent. More importantly, once it becomes self-evident that the exponents measured at different scales have to differ because of how arithmetic operations are applied, the real question  becomes whether this simple  and \emph{necessary} disparity is in any way biologically meaningful. Here the saturation exponents observed at regional scales potentially tells us nothing about how yield, or how other ecosystem properties change with regional diversity. Other than being an arithmetic property of exponentiation, as far as we can tell, this regional exponent provides no meaningful biological information.

It is important to understand that if instead of the \emph{observed} regional exponent $\beta^*_{\mathcal{M}}$, we used $\beta_{\mathcal{M}}$ as an explicit function of the underlying metacommunity variables, the derivative $\frac{\mathrm{d}}{\mathrm{d}N} \; y_{_\mathcal{M}}$ in Eq. \eqref{eq:der_y_metacomm} would have equalled the derivative $\frac{\mathrm{d}}{\mathrm{d}N} \sum_j y_{_{\mathrm{Total},j}}$. Given that both the total yield at the metacommunity, and the yield summed across sites have to equal each other, $y_{_\mathcal{M}} = \sum_j y_{_{\mathrm{Total},j}}$, the derivatives of both functions must also, by definition, be equal when the the regional exponent $\beta^*_{\mathcal{M}}$ is explicitly considered as a function of underlying variables, as seen in Eq. \eqref{eq:beta_M} (this would mean applying the chain rule in Eq. \eqref{eq:der_y_metacomm} to explicitly differentiate $\beta_{\mathcal{M}}$ wrt to $N$: $\frac{\textrm{d}\beta_{\mathcal{M}}}{\textrm{d}N}= \frac{\textrm{d}\beta_{\mathcal{M}}}{\textrm{d}\beta} \; \frac{\textrm{d}\beta}{\textrm{d}N} $). But doing so would mean acknowledging that the exponent $\beta^*_{\mathcal{M}}$ is a pure \emph{mathematical construct} of other, more meaningful parameters.

If the regional exponent only makes sense if one already knows all the details of the underlying structure of the metacommunity, along with all the parameters and variables associated with that structure, then the values  of the observed exponent, $\beta^*_{\mathcal{M}}$, will be redundant since one still needs knowledge of all the local parameters to determine the regional scale saturation of yield; the exponent $\beta^*_{\mathcal{M}}$ on its own tells us nothing. In fact, pretending that the observed values of $\beta^*_{\mathcal{M}}$ represent some form of biological phenomena  itself represents the height of reification -- that is, the error of ``thingifying'' a purely abstract or mathematical construct.

The value represented by $\beta^*_{\mathcal{M}}$ appears to be nothing more than a mathematical contrivance -- an artefact of the non-additivity of exponential functions (which Thompson et al., 2018, incorrectly found to be additive). Without any meaningful biological interpretation, such parameters run the risk of being either, at best, redundant, or worse, outright misleading -- or even worse still, a form of mystification.

\section{Discussion and Conclusion}

We have outlined in this paper various pseudo-biodiversity effects that can arise when measuring so-called ecosystem functioning across both space \citep{qiu.cardinale_20}, and over time \citep{reich2012,huang.chen.ea_18,qiu.cardinale_20}. The first pseudo effect discussed, Population-level effects, can arise when considering the temporal changes that occur in a single species population in the absence of any biodiversity. The other two effects -- Independence and Arithmetic effects -- can be seen when quantifying properties over increasing spatial scales. These latter two effects, in particular, can be considered as a measure of the degree to which some property of interest is trivially non-intensive (see discussion below).

There are a couple of points that one needs to keep in mind when considering the set of arguments outlined in this paper. First, the term``trivial'' here does not mean that the quantities obtained when measuring biodiversity effects across scales are false, or that they are being measured incorrectly.  Rather, triviality  arises from the fact that the real quantities measured are either artefacts of how one counts, quantifies or classifies a set of entities or their attributes (whether these entities are biological or not); or triviality arises as a simple consequence when one measures the obvious, and otherwise unremarkable biological features of a system -- like growth in a single population, or the non homogenous nature of the biosphere across large spatial scales. Neither of these artefactual features of quantification can tell us anything meaningful about role of biodiversity, however conceived.  

Secondly, it should be emphasized that our claim is not that \emph{all} measured quantities are necessarily trivial, just that there is a trivial, non-biological component built in to quantities obtained when measuring ecosystem properties across scales. In order to illustrate the presence of this trivial component the theoretical exposition of this paper required that we make a set of highly idealized assumptions in order to remove any real biological effects that could be confounded with the trivial non biological effects that we were trying to demonstrate as always operating in biodiversity studies (i.e., Assumptions \ref{a1}-\ref{a6}). All of our examples were set up to demonstrate how standard BEF approaches would detect ecosystem functioning even when we ensured that no real biodiversity effects operated -- either by only allowing effects that arose from growth of single population (Population-level effects), or non biological effects that are artefacts of how one counts or measures across spatial scales (Independence and Arithmetic effects). 

Recall that when considering spatial scales, every site in our study was identical in ecosystem functioning or yield, and that all species were equivalent within and between sites, establishing a hypothetical idealized system where ecosystem functioning could be considered independent of biodiversity.  Thus we constructed a landscape that was uniform and homogenous with regard to functioning/yield, where no matter how many species we added to the system, we would see no increase in ecosystem functioning at a given scale. What all species being effectively indistinguishable both within and between sites meant was that the artificial net biodiversity effect that arose from Independence effects was also in part an artefact of how species are arbitrarily classified or the degree of taxonomic resolution or reclassification. To take pseudo  effects that arise from changes in OTU resolution seriously and without qualification as being a type of biodiversity effect would require us accepting absurd constructs, such as the concept of a "metabiome", where the measurement of taxonomic turnover -- even across  geographic biomes like marine and terrestrial systems, or across Arctic and tropical zones -- could be construed as a meaningful biodiversity effect.

Arithmetic effects represent an even more striking case of non-meaningful quantification, particularly as seen in studies where the behaviour of mathematical functions or operations are misconstrued as being the result of natural phenomena \citep{thompson.isbell.ea_18}. This type of pseudo quantification (particularly in the form of simulation-based studies) is analogous to another problematic form of study in ecology: the use of experiments to test the validity of analytic or mathematical statements \citep[see][]{gouhier.pillai_19a} . The use of simulation and experimental approaches in this way comes as close as one can come to the inappropriate use of empirical work to `test' the truth value of analytic propositions, such as attempting to experimentally validate whether ``$5+7=12$''. Pseudo quantification and pseudo empirical studies are likely to be far more prevalent in ecological research than currently acknowledged \citep{gouhier.pillai_19a}.

The examples of artefactual effects we demonstrated in this paper for Independence and Arithmetic effects can be framed more abstractly in terms of the nature of measured properties (ecosystem functions) being either intensive or non-intensive. Intensive properties or quantities are simply those who's magnitudes do not change with the size of the system (independent of system size), as in the case of physical quantities like density or temperature. Behind much of the recent interest in measuring ecosystem properties across spatial scales is the implicit belief that if the quantification of some ecosystem effect or property is found to be non intensive in nature then any observed effects arising from changes in the spatial scale of the system must be meaningful. Given any large scale system of size $m$ (representing, for example, the number of sites in a metacommunity), the magnitude of the system-level (intensive) property $\Phi$ can be expressed as a function of the property of its smaller subunits, $\varphi$ in the following manner,
\begin{align}
  \Phi \left(\;m \cdot \varphi \; \right) \; &= \; \Phi\left(\varphi\right) \;\;\;\;\; (\text{Intensive property}).
  \end{align}
(Note how the total property of the system is independent of its size, $m$, and only depends on the average magnitude of the subunits.) 

Potentially two types of trivial effects can lead to measured quantities being non-intensive at large spatial scales. The simplest and most obvious arises when system-level properties (such as biomass) are \emph{extensive} and thus add or scale up linearly as quantities measured in local sites are aggregated (i.e., the properties of smaller subunits/sites are additive),
\begin{align}
  \Phi \left(m \cdot \varphi \right) \; &= \; m\cdot \Phi\left(\varphi\right) \;\;\;\; (\text{Extensive property}).
\end{align}
This form of triviality arising from a linear dependence on system size is also the most easy to control for (as was done in Eqs. \eqref{eq:Y_scaled}-\eqref{eq:Delta_scaled}). Because the ratio of any two extensive properties is itself intensive we can control for this type of triviality by dividing a total system-level property by its size (Eqs. \eqref{eq:Y_scaled}-\eqref{eq:Delta_scaled}; Fig.\ref{Fig:independence_effects}). As we would  expect, the raw biodiversity effect measured as total yield, $\sum_j y_{_\mathrm{Total},\;j}$ is an extensive property, that when scaled by $m$ gave rise to an intensive quantity, $1/m \sum_j y_{_\mathrm{Total},\;j}$, that was unaffected by system size (horizontal line Fig. \ref{Fig:independence_effects}b). On the other hand, the net biodiversity effect at the regional scale was not extensive and scaling it by $m$ still resulted in a monotonically increasing quantity (Fig. \ref{Fig:independence_effects}d). However, just because a non intensive property (like the net biodiversity effect at the regional scale) is also non extensive in nature does not mean it represents a meaningful quantity.

For example, in addition to triviality being due to properties being extensive, another less obvious type of triviality can arise when the effects are either artefacts of turnover and how species/OTUs are counted and categorized, or effects whose magnitudes are simply artefacts of how mathematical functions behave. These  pseudo effects represent additional non-biological, or at least non-meaningful ecological effects whose role we were able to demonstrate deductively. The regional net biodiversity effect, $\Delta_{_\mathrm{Net}}^\mathcal{M}$, and the regional-scale exponent parameter, $\beta^*_\mathcal{M}$ (Eq. \eqref{eq:beta_measured}), were two quantities that were shown to be neither intensive nor extensive properties, yet they still represented effects that were biologically non-meaningful in nature. 

It is also useful to note that our classification of the different pseudo biodiversity effects was not meant to represent a set of mutually exclusive categories. In fact, certain pseudo effects may only arise in the context of other effects, as in cases where arithmetic effects only become apparent when independence effects operate. In our example (the metacommunity described in Eq. \eqref{eq:saturation_metacomm}), Independence effects being absent would imply no species turnover between sites; with no turnover the regional scale saturation exponent parameter will equal the value of the local exponent, $\beta_\mathcal{M}^*=\beta$, and thus Arithmetic effects will vanish (see Eq. \eqref{eq:Appendix_beta_M_generalexpression} in Appendix \ref{saturationexponent}). Similarly, OTU reclassification through aggregation (or disaggregation) can potentially decrease (or increase) the pseudo quantification of biodiversity effects that fall under Arithmetic effects. 

\subsection*{Concluding remarks}
This study was not an attempt at modelling specific or general ecological processes, but an analytical demonstration of how, under very simple assumptions, pseudo-biodiversity effects could be confounded with true biodiversity effects in natural experiments. The fact that little or no real biological features are needed to demonstrate the presence of artefactual effects in BEF studies carried out across scales (both in experiments and models) only serves to demonstrate how the results obtained from such studies may represent little more than the built-in properties of how certain rudimentary arithmetic and mathematical functions behave.




\bibliographystyle{ecology_letters}
\bibliography{pillai_pseudo_biodiversity_effects_2020.bib}

\newpage{}
\section*{Appendices}

\setcounter{subsection}{0} \renewcommand{\thesubsection}{A\arabic{subsection}}

\setcounter{equation}{0} \renewcommand{\theequation}{\thesubsection.\arabic{equation}}
\setcounter{figure}{0} \renewcommand{\thefigure}{\thesubsection.\arabic{figure}}

\subsection{Independence effects arising from OTU resolution} \label{otu_resolution}
Calculations of $\Delta^{^\mathcal{M}}_{\mathrm{Net}}$ at different OTU resolutions can be carried out using the simple equation $\Delta^{^\mathcal{M}}_{\mathrm{Net}}= \left( m-\overline{\nu}\right)Y$, as was done in the main text (here $\overline{\nu}$ represents the average number of sites an OTU occupies in the metacommunity). To derive the expression for this formula we first start with the following equation for an $m$-site metacommunity with a regional diversity of $N$,
\begin{align} \label{eq:Appendix_num_slots_expression_a}
  m \; \overline{n} &= N \; \overline{\nu} , 
\end{align}
where  $\overline{n}$ is the average within-site OTU diversity, and $\overline{\nu}$ is the average OTU site-occupancy (see Appendix \ref{saturationexponent} for details). 

Recall from Eq. \eqref{eq:metacomm_null} that the expected biodiversity effect under our previous set of assumptions will be 
\begin{align} 
  \text{Expected (LH baseline)} &= \frac{1}{mn} \sum_{j\in \mathcal{M}} \sum_{i=1}^n Y_{i,j}  \notag \\
  &= Y. 
\end{align}
Now, if we no longer assume that the number of species/OTUs in all sites are the same, and if we no longer assume that each species/OTU will only appear in a single site, then we will get the following expression for the LH or expected baseline:
\begin{align} 
  \text{Expected (LH baseline)} &= \frac{1}{N} \sum_{j\in \mathcal{M}} \sum_{i=1}^n Y_{i,j}  \notag \\
  &= \frac{1}{N} \times Y \sum_{j\in \mathcal{M}} n_j  \notag \\
  &= \frac{1}{N} \times m \; \overline{n} \; Y .   
\end{align}
Combining this new expression with Eq. \eqref{eq:Appendix_num_slots_expression_a} above gives the formula for the LH null, 
\begin{align} 
  \text{Expected (LH baseline)} &= \overline{\nu} \; Y.   
\end{align}
If we return to the equation for the net biodiversity effect at the metacommunity scale, $\Delta^{^\mathcal{M}}_{\mathrm{Net}} =\text{Observed} - \text{LH null}$ (as seen in Eq. \eqref{eq:net_at_metacomm}), and replace the original expression for the LH null with our new version, we then obtain the desired formula,   
\begin{align} 
  \Delta^{^\mathcal{M}}_{\mathrm{Net}} &=\text{Observed} - \text{LH null}  \notag \\
  &= \left(m -\overline{\nu}\right) Y.  
\end{align}

\begin{figure}[h!]
  \centering
  \includegraphics[width=0.65\textheight]{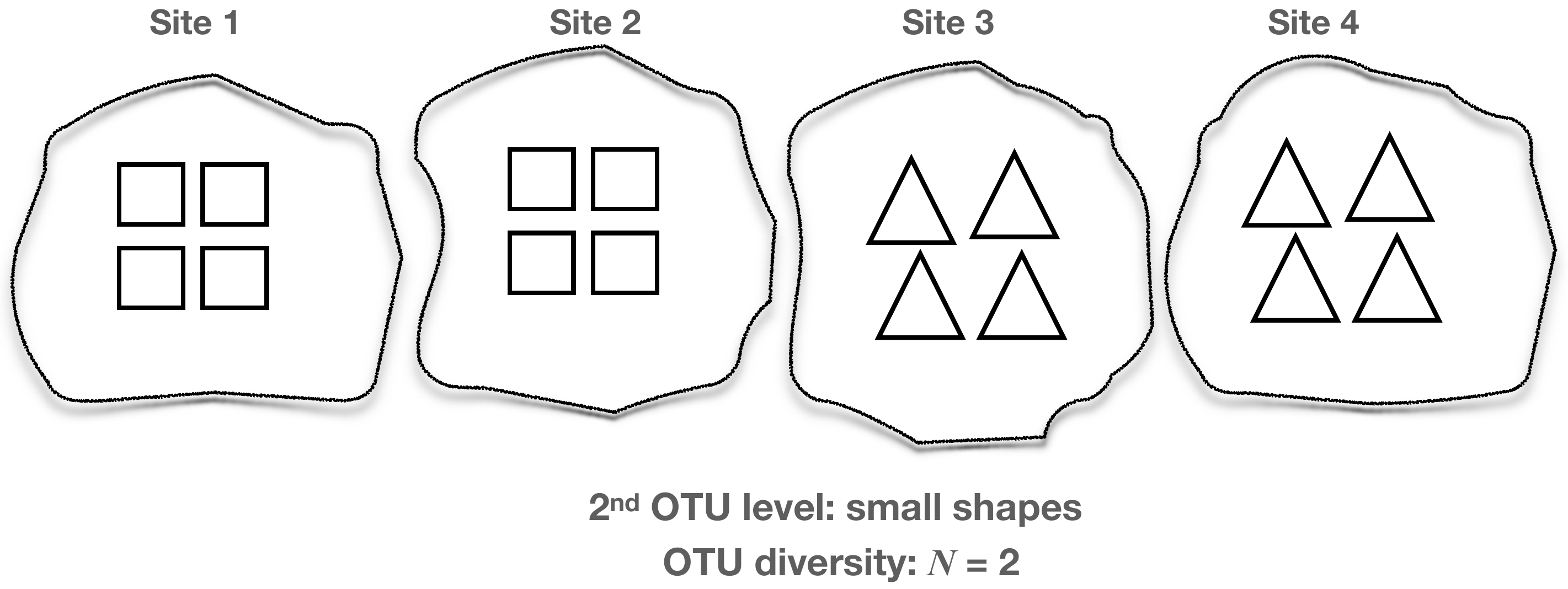}
  \caption{ $2^{nd}$ level OTU resolution.}{}
  \label{Fig:OTU_2nd}
\end{figure}
We can briefly demonstrate here how to apply this formula to the metacommunity example illustrated in Fig. \ref{Fig:OTU_resolution} in the main paper. For the second OTU level (Fig. \ref{Fig:OTU_2nd} ) we can see that there are two OTUs (squares and triangles) in a metacommunity with $m=4$ sites, and where the average OTU site-occupancy is $\overline{\nu}=2$. Thus the net biodiversity effect at the metacommunity scale will be $\Delta^{^\mathcal{M}}_{\mathrm{Net}}= \left( m-\overline{\nu}\right)Y = 2Y$.

\begin{figure}[h!]
  \centering
  \includegraphics[width=0.65\textheight]{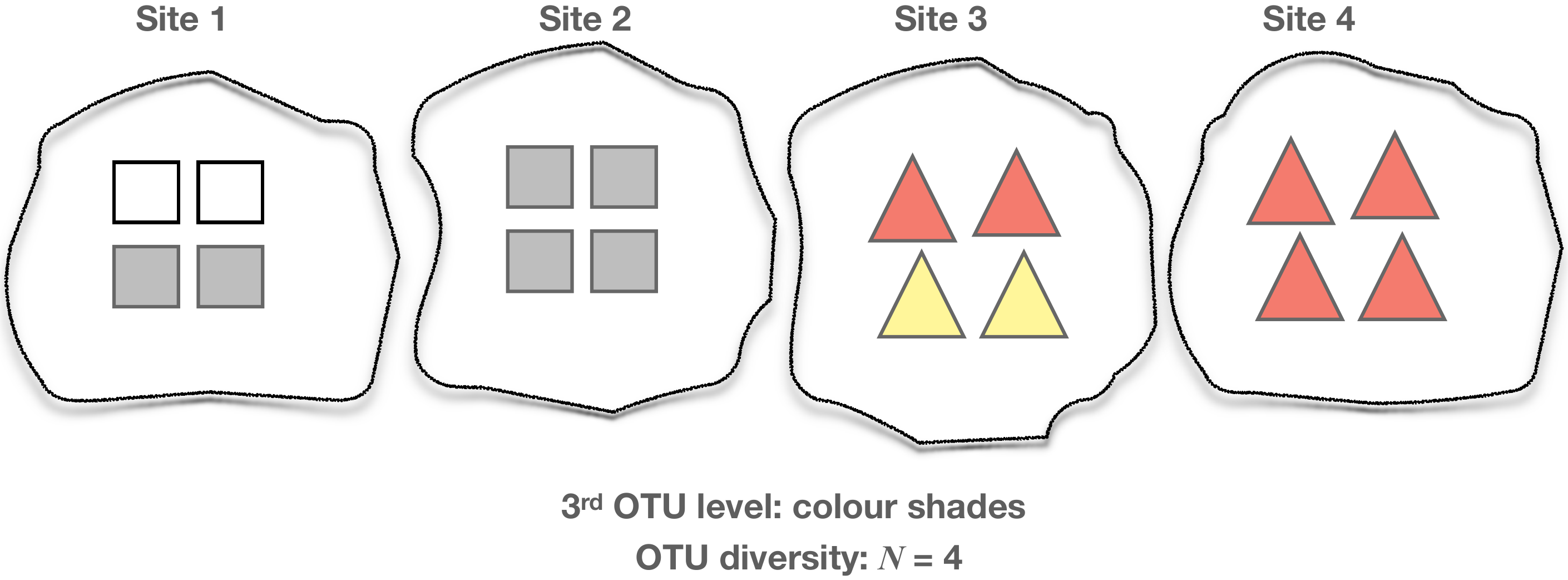}
  \caption{ $3^{rd}$ level OTU resolution.}{}
  \label{Fig:OTU_3rd}
\end{figure}

Similarly for the third OTU level (Fig. \ref{Fig:OTU_3rd} ) we have a regional diversity of $N=4$ (four colour shades: white, grey, yellow, and red), with an average site-occupancy of $\overline{\nu}=1.5$, which gives a net biodiversity effect of $\Delta^{^\mathcal{M}}_{\mathrm{Net}}= 2.5Y$.

\setcounter{equation}{0} \renewcommand{\theequation}{\thesubsection.\arabic{equation}}
\setcounter{figure}{0} \renewcommand{\thefigure}{\thesubsection.\arabic{figure}}

\subsection{General expression for independence effect} \label{independence_general}
In order to generalize the net biodiversity effect at the metacommunity scale we will relax some key assumptions made previously: specifically, that all sites have an equal number of species (Assumption \ref{a1}); that all species within a site have the same monoculture yields, and that these monoculture yields are the same in all sites (Assumption \ref{a3}); and finally, that all species only occupy a single site in the metacommunity (Assumption \ref{a6}).

Given the above, the expression for the LH null at the regional scale for $N$ species across $m$ sites will be given by 
\begin{align}
  \renewcommand{\arraystretch}{0.7}\begin{array}{c} \text{LH null} \\ \text{(metacommunity-scale)} \end{array} &= \; \frac{1}{N} \sum_i^N \sum_j^m Y_{i,j}, \notag \\
  &=  \; \frac{1}{N} \sum_i^N \nu_i \; \langle Y_{\text{sp }i} \rangle , \notag \;\;\;\; \notag  
\end{align} 
Here $\nu_i$ is the number of sites occupied by species $i$ and $\langle Y_{\text{sp }i} \rangle$ is average monoculture yield of species $i$ across all the sites it occupies. If we take $\mathcal{N}$ to be the random variable for occupancy rate $\nu$ and $\langle \mathcal{Y_{\mathrm{sp}}} \rangle $ to represent the random variable for $\langle Y_{\text{sp }i} \rangle$, then the LH null can be written out as 
\begin{align}
  \renewcommand{\arraystretch}{0.7}\begin{array}{c} \text{LH null} \\ \text{(metacommunity-scale)} \end{array} &= \; \mathbb{E}\Bigl[ \mathcal{N} \cdot \langle \mathcal{Y_{\mathrm{sp}}} \rangle   \Bigr]  \notag \\
  &=  \; \overline{\nu} \cdot\overline{\langle Y_{\mathrm{sp}} \rangle} \;+\; \mathrm{cov} \Bigl[ \mathcal{N} , \langle \mathcal{Y_{\mathrm{sp}}} \rangle \Bigr] .  \;\;\;\; \label{eq:LH_null_regional} 
\end{align} 

If we take the total observed biodiversity effect across the region to be $\sum_j^m y_{_{\mathrm{obs},j}}$, then the regional or metacommunity net biodiversity effect $\Delta_{_\mathrm{Net}}^\mathcal{M}$ will simply be the total observed minus the regional LH null:
\begin{align}
  \Delta_{_\mathrm{Net}}^\mathcal{M} &= \; \sum_j^m y_{_{\mathrm{obs},\; j}} \;- \; \overline{\nu} \cdot\overline{\langle Y_{\mathrm{sp}} \rangle} \;-\; \mathrm{cov} \Bigl[ \mathcal{N} , \langle \mathcal{Y_{\mathrm{sp}}} \rangle \Bigr]. 
\end{align}

Accounting for the sum of all \emph{local} LH nulls within sites can allow us to separate out a trivial component of this regional net biodiversity effect, 
\begin{align}
  \Delta_{_\mathrm{Net}}^\mathcal{M} &=  \Biggl( \sum_j^m y_{_{\mathrm{obs},\; j}} - \; \left(\substack{\text{aggregate}\\ \text{of local} \\ \text{LH nulls} } \right)  \Biggr) + \Biggl( \left(\substack{\text{aggregate}\\ \text{of local} \\ \text{LH nulls}} \right)- \overline{\nu} \cdot\overline{\langle Y_{\mathrm{sp}} \rangle} \;-\; \mathrm{cov} \Bigl[ \mathcal{N} , \langle \mathcal{Y_{\mathrm{sp}}} \rangle \Bigr] \Biggr). 
\end{align}
Assuming the initial proportion of each species in a site is simply 1 divided by the number of species in that site, $ 1 /n_j$, then the aggregate of the local LH nulls across all sites will be 
\begin{align}
  \renewcommand{\arraystretch}{0.7}\begin{array}{c} \text{Local LH nulls} \\ \text{(aggregated)} \end{array} &= \;  \sum_j^m \sum_i^{n_j} \frac{1}{n_j} Y_{i,j} \notag \\
  &= \; \sum_i^m \langle Y_{\mathrm{site}\; j} \rangle \notag \\
  &= \; m \cdot \overline{ \langle Y_{\mathrm{site } \; j} \rangle },
\end{align}
which leads to an expression of the net biodiversity effect as the sum of its trivial and less trivial components: 
\begin{align}
  \Delta_{_\mathrm{Net}}^\mathcal{M} &= \color{red}\underbrace{\color{black} \left(\sum_j^m y_{_{\mathrm{obs},j}} - m\cdot\overline{\langle Y_\mathrm{site} \rangle} \; \vphantom{\sum_j^m} \right) }_{\substack{ \text{Less trivial component} } } \color{black} \;+\; \color{blue}\underbrace{\color{black} \; \left( m\cdot\overline{\langle Y_{\mathrm{site}} \rangle} - \overline{\nu} \cdot\overline{\langle Y_{\mathrm{sp}} \rangle} - \mathrm{cov}\Bigl[ \mathcal{N} , \langle \mathcal{Y_{\mathrm{sp}}} \rangle \Bigr] \vphantom{\sum_j^m} \right)}_{\substack{ \text{Trivial component} \\ \text{(effect of turnover)} } } .  \label{eq:general_net} 
\end{align}

When the monoculture yields for a given species is the same in each site it occupies (i.e., $Y_{i,j}=Y_i$), then we have $\langle Y_{\mathrm{sp}} \rangle = Y_i$. Furthermore, if all species have the same monoculture yields in all sites then we get $\langle Y_{\mathrm{site}} \rangle = Y$ and $\langle Y_{\mathrm{sp}} \rangle= Y$. Making these simplifications (as per Assumption \ref{a3}) leads to
\begin{align}
  \overline{\langle Y_{\mathrm{sp}} \rangle} &=  \; \frac{1}{N} \sum_i^N \sum_j^m Y_{i,j} =  \frac{1}{N} \sum_i^N \nu \; \langle Y_{\mathrm{sp}} \rangle = \overline{\nu} \; Y, 
\end{align}
and 
\begin{align}
  \overline{\langle Y_{\mathrm{site}} \rangle} &=  \; \frac{1}{m} \sum_j^m \sum_i^{n_j} \frac{1}{n_j} Y_{i,j} =  \frac{1}{m} \sum_i^m \langle Y_{\mathrm{site}} \rangle = Y. 
\end{align}
With these assumptions the trivial component in Eq. \eqref{eq:general_net} becomes
\begin{align}
  \Delta_{_\mathrm{Trivial}} &=    (m-\overline{\nu}) \; Y - \cancelto{\;0}{ \mathrm{cov}\Bigl[ \mathcal{N} , \langle \mathcal{Y_{\mathrm{sp}}} \rangle \Bigr] }. 
\end{align}
If we also assume perfect turnover (as per Assumption \ref{a6}), then the average site occupancy for all species becomes $\overline{\nu} = 1$, and the trivial component of the metacommunity net biodiversity effect reduces to
\begin{align}
  \Delta_{_\mathrm{Trivial}} &=    (m-1) \; Y , 
\end{align}
as we saw in Eq. \eqref{eq:net_at_metacomm} in the main paper.

\setcounter{equation}{0} \renewcommand{\theequation}{\thesubsection.\arabic{equation}}
\setcounter{figure}{0} \renewcommand{\thefigure}{\thesubsection.\arabic{figure}}

\subsection{Solving for the saturation curve exponent} \label{saturationexponent}

In the main paper we used $Y_{i,j}$ as the coefficient of species $i$'s monoculture property (e.g., yield) in site $j$. Here we will replace the monoculture yield value  $Y_{i,j}$ with the general coefficient $a_{i,j}$ . For simplicity all species are assumed in a given site to have identical property/yield monocultures, so $a_{i,j}$ is a constant $a_j =a$. The total property/yield of a multispecies community in site $j$ will scale with the  local diversity level, $n_j$,  by $a_j \times n_j^{\beta_j}$ (for exponent $\beta_j$). Summing across all sites in a metacommunity (defined by the index set of sites $\mathcal{M}$) gives the metacommunity or regional scale version of the saturation curve,
\begin{align} \label{eq:Appendix_saturation_metacomm}
   a_{_\mathcal{M} } \; N^{{ \beta_{\mathcal{M}}} }  &= \sum_{j\in\mathcal{M}} a \; n_j^{\beta_j}, 
\end{align}
where the metacommunity diversity $N$ is the sum of all local diversity levels, $N =\sum_j n_j$, and $a_\mathcal{M}$ is the property/yield coefficient of the average species at the metacommunity scale given by the index set $\mathcal{M} = \{1, \ldots, m \}$.

The exponent $\beta_{\mathcal{M}}$ is a parameter measured at the regional metacommunity scale. What should be obviously clear is that this metacommunity-scale exponent \emph{has} to be different from each of the exponents in the individual sites, $\beta_j$ -- except in the case where $\beta_j$ in each site is equal to 1. There is thus no need to test whether the exponents at the local and regional scales differ, as was done in \citet{thompson.isbell.ea_18}. Furthermore, if one wanted to find out \emph{how} the exponents differ (assuming for the moment that this would represent a meaningful or appropriate research question) one could have just simply written out the expression defining the relationship between the two \emph{explicitly}, instead of running simulations to generate values that could then be subject to a linear regression. Yet the whole objective and methodological approach of \citet{thompson.isbell.ea_18} was precisely to explore the relationship between the exponent of a sum and the sum of exponents via simulations.

For sake of completion we will in this appendix section write out the explicit formula showing the relationship between sum of exponents and the exponentiation of a sum  (using the approach one would find in any elementary arithmetic textbook).

The first thing to note is that the authors of \citet{thompson.isbell.ea_18} either did not realize that the metacommunity coefficient $a_{_\mathcal{M}}$ will not necessarily be the same as the coefficient $a$ found in each site, or worse did not think it an issue to resolve. To continue with the approach of the authors of \citet{thompson.isbell.ea_18} we will maintain that the coefficient at the metacommunity scale, $a_{_\mathcal{M}}$, provides a representative average of all species' total metacommunity monoculture yields taken together. For the moment we will assume the simplest scenario, specifically: that all sites have identical diversity levels, $n$; that there is perfect species turnover between sites such that each species appears in only one site; and finally, that all sites have the same saturation exponent such that $\beta_j =\beta$. Under these simplifying assumptions the metacommunity coefficient $a_{_\mathcal{M}}$ will be equal to the total metacommunity-scale monoculture yields of each species, $a_{_\mathcal{M}} =  a$.

Using the above assumptions if we simply log  both sides of Eq. \eqref{eq:Appendix_saturation_metacomm} and rearrange the terms we get the following expression for a metacommunity with $m$ sites:
\begin{align} \label{eq:beta_M_b}
   \beta_{_\mathcal{M} } &= \frac{\beta \log n \;+ \log m \; }{ \;\;\;\log n \;+ \log m \;}  \;\;\;\;\;\; (\text{for} \;\;\; m\times n \neq 1). 
\end{align}
This straightforward expression belies any need for running simulations to demonstrate the relationship between local and regional exponents. Furthermore, simple inspection of the above expression shows that the exponent applied to a sum of terms, $\beta_{_\mathcal{M} }$,  will \emph{not} equal the exponent that was applied to the individual terms (bases) in the sum, $\beta$ (except when $\beta=1$).

If we wish to relax some of the above assumptions regarding  species turnover, diversity level and the values of the exponent $\beta_j$  in each site we will need to define some new parameters. First we will define the degree of species `turnover', $\mu$, as the regional species pool divided by the average local site diversity:
\begin{align} \label{eq:Appendix_turnover}
   \mu & \doteq  \frac{N}{\overline{n}}  \;\;\;\;\;\; (\text{where} \;\;\; 1\leq \mu \leq m). 
\end{align}
Perfect turnover occurs when $\mu=m$, that is where each species occurs in only one site  and all sites are completely distinct in  regards to species composition. No turnover occurs when $\mu=1$ and all sites are identical and each site contains the regional species pool.

Now imagine $m$ sites with the diversity level in each $j$ site as $n_j$. Consider the diversity level in a site as the number of species slots to be filled in the site. Then the total number of species habitat slots to fill in the metacommunity is
\begin{align} \label{eq:Appendix_num_slots}
   n_1 + n_2 +  \ldots + n_m &= \sum_{j=1}^m n_j    \notag \\
   &= m \; \overline{n}
\end{align}

For a species $i$ is  in the metacommunity species pool, (where $1 \leq i \leq N$),  $\nu_i$ will represent the `site incidence', or number of sites in the metacommunity where species $i$ appears ($1 \leq \nu_i \leq m$). The total number of species slots to fill in the metacommunity then is given by
\begin{align} \label{eq:Appendix_num_slots_nu}
   \nu_1 + \nu_2 +  \ldots + \nu_N &= \sum_{i=1}^N \nu_i   \notag \\
   &= N \; \overline{\nu}
\end{align}
Since Eq. \eqref{eq:Appendix_num_slots} and Eq. \eqref{eq:Appendix_num_slots_nu} both represent the number of species slots to be filled they must be equal to each other, which gives us
\begin{align} \label{eq:Appendix_num_slots_expression}
   m \; \overline{n} &= N \; \overline{\nu} . 
\end{align}
Combining Eq. \eqref{eq:Appendix_num_slots_expression} with the definition of turnover in Eq \eqref{eq:Appendix_turnover} then provides us with a useful expression of average species incidence or site occupancy in the metacommunity,
\begin{align} \label{eq:Appendix_incidence}
   \overline{\nu} &= \frac{m}{\mu}. 
\end{align}

The biodiversity-ecosystem saturation curve at the regional scale is represented by defining the metacommunity coefficient $a_{_\mathcal{M}}$ as the representative average of the total metacommunity monoculture yields of all species taken together. We will assume that that the local monoculture yields each species in a given site (where they exist), $a_{i,j}$ will be the same for all species in any of the sites they occupy, such that $a_{i,j} =a$. This gives us
\begin{align} \label{eq:Appendix_a_M}
   a_{_\mathcal{M}} &= \frac{1}{N} \sum_{i=1}^N \sum_{j=1}^m a_{i,j}   \notag \\
   &= \frac{1}{N} \times \sum_{i=1}^N a_i\times\nu_i  \notag \\
   &= \frac{1}{N} \times a\sum_{i=1}^N \nu_i  \notag \\
   &= a \; \overline{\nu}.
\end{align}

Returning to the expression for the metacommunity scale saturation curve $a_{_\mathcal{M} } \; N^{{ \beta_{\mathcal{M}}} }$ (the LHS of Eq. \eqref{eq:Appendix_saturation_metacomm} ) we  can incorporate the above expressions to rewrite the expression as
\begin{align} \label{eq:Appendix_metacomm_exponential}
   a_{_\mathcal{M} } \; N^{{ \beta_{\mathcal{M}}} } &= a\;\overline{\nu} \times (\mu\; \overline{n})^{{ \beta_{\mathcal{M}}} } = a\;\frac{m}{\mu} \times (\mu\; \overline{n})^{{ \beta_{\mathcal{M}}} }.  
\end{align}

For the simple case where the diversity level and saturation exponent in each site is the same, $n_j=n$ and $\beta_j =\beta$, we can  rewrite  \eqref{eq:Appendix_saturation_metacomm} by replacing the LHS with the expression for $a_{_\mathcal{M} } \; N^{{ \beta_{\mathcal{M}}} }$ obtained in Eq. \eqref{eq:Appendix_metacomm_exponential},  and then log both sides to obtain an  explicit expression for the metacommunity saturation exponent,  $\beta_{\mathcal{M}}$, in terms of the  local exponents:
\begin{align} \label{eq:Appendix_saturation_metacomm_explicit}
   a_{_\mathcal{M} } \; N^{{ \beta_{\mathcal{M}}} }  &= \sum_{j\in\mathcal{M}}^m a \; n_j^{\beta_j} \notag \\
   a\;\frac{m}{\mu} (\mu\; n)^{{ \beta_{\mathcal{M}}} }  &= m \; a \; n^{\beta}  \notag \\
   {{ \beta_{\mathcal{M}}} } \log(\mu\; n)  &= \beta \log(n) + \log(\mu) 
\end{align}
This gives
\begin{align} \label{eq:Appendix_beta_M_generalexpression}
   \beta_{_\mathcal{M} } &= \frac{\beta \log n \;+ \log \mu \; }{ \;\;\;\log n \;+ \log \mu \;}  \;\;\;\;\;\; (\text{for} \;\;\; \mu\times n \neq 1). 
\end{align}
Since $1\leq\mu\leq m$, this means the above expression holds for when both $m\neq1$ and $m\neq1$.

When turnover is complete, and all sites contain different species ($\mu = m$), then Eq. \eqref{eq:Appendix_beta_M_generalexpression} becomes the expression described in the main paper (see Eq.\eqref{eq:beta_M} )
\begin{align} \notag
   \beta_{_\mathcal{M} } &= \frac{\beta \log n \;+ \log m \; }{ \;\;\;\log n \;+ \log m \;}  \;\;\;\;\;\; (\text{for} \;\;\; m\times n \neq 1). 
\end{align}
If on  the other hand there is no turnover and all sites are identical ($\mu = 1$), then the above expression reduces to $\beta_{_\mathcal{M} } = \beta$, which is unsurprising since in this case the metacommunity is effectively a single site. Thus, only when $\mu = 1$ or when $\beta=1$ will the relationship $\beta_{\mathcal{M}} =\beta$ hold, contrary to the simulation results of \citet{thompson.isbell.ea_18}.

In general, the other cases considered by authors of \citet{thompson.isbell.ea_18} can also be analysed by writing out the general expression for the metacommunity exponent, $\beta_{_\mathcal{M} }$, explicitly (see Appendix \ref{saturationecurve} for details). Following this straightforward approach can allow  one to easily demonstrate that the other key results presented in  \citet{thompson.isbell.ea_18} cannot hold.

\setcounter{equation}{0} \renewcommand{\theequation}{\thesubsection.\arabic{equation}}
\setcounter{figure}{0} \renewcommand{\thefigure}{\thesubsection.\arabic{figure}}

\subsection{Effect of increasing diversity on the metacommunity saturation curve} \label{saturationecurve}

\subsubsection{Simple expressions for the metacommunity saturation exponent}
The metacommunity biomass  of species $i$ in site $j$ is represented by the monoculture coefficient  $Y_{i,j}$.

Now, as in the  Thompson et al (2018) paper, we will consider how a property like total yield in the metacommunity, $ y_{_{\mathcal{M}}  } = \sum_{j\in\mathcal{M}} y_{_{\mathrm{Total},j} }$, will saturate with increasing biodiversity, and how the shape of the saturation curve will differ depending on whether the metacommunity is treated either as a single site, $y_{_{\mathcal{M}}  }= Y_{_{\mathcal{M}}} N^{{ \beta^*_{\mathcal{M}}} }$,  or as a series of sites each with their own individual saturation functions, $\sum_{j\in\mathcal{M}} y_{_{\mathrm{Total},j} } = \sum_{j\in\mathcal{M}} Y_{i,j} \; n_j^{\beta_j}$, where ${{ \beta^*_{\mathcal{M}}} }$ is a purported saturation exponent observed at the regional (metacommunity) scale,
\begin{align} \label{eq:Appendix_regional_biomass}
   y_{_{\mathcal{M}}  }  &= \sum_{j=1}^m y_{_{\mathrm{Total},j} }  \notag \\
   Y_{_{\mathcal{M}}} N^{{ \beta^*_{\mathcal{M}}} } &= \sum_{j=1}^m Y_{i,j} \; n_j^{\beta_j}. 
\end{align}
Solving the above for for ${{ \beta^*_{\mathcal{M}}} }$ gives the general expression for the saturation exponent:
\begin{align}
   {{ \beta^*_{\mathcal{M}}} }  &= \frac{ \log(\sum_{j=1}^m y_{_{\mathrm{Total},j} })\; - \log(Y_{_{\mathcal{M}}}) \; }{ \;\;\;\log N \;\;}  \notag \\
   &= \frac{ \log(\sum_{j=1}^m Y_{i,j} \; n_j^{\beta_j})\; - \log(Y_{_{\mathcal{M}}}) \; }{ \;\;\;\log N \;\;}  \;\;\;\;\;\; (\text{for} \;\;\; N \neq 1), 
\end{align}

We can simplify the expression further if we assume that the biomass coefficient of each species $i$ is constant regardless  of the site it occupies (so long as it is exists in the site):
\begin{align}
    Y_{i,j} &=
\begin{dcases}
    Y_i ,& \text{if species $i$ can exist in site $j$}  \\
    0,              & \text{if species $i$ cannot exist in site $j$}
\end{dcases} &{}
\end{align}
Furthermore, we will assume that all species have identical monoculture coefficients  to each other, such that $Y_i =Y$. Making the above assumptions will allow us to use the species site occupancy rate $\nu$ (from Appendix A1) to rewrite the coefficient $Y_{_{\mathcal{M}}}$ in the following manner: $ Y_{_{\mathcal{M}}}= \frac{1}{N} \sum_i^N \nu_i Y_i = \overline{\nu}\; Y$. This yields
\begin{align}
   {{ \beta^*_{\mathcal{M}}} }  &= \frac{ \log(\sum_{j=1}^m  n_j^{\beta_j})\; - \log(\overline{\nu}) \; }{ \;\;\;\log N \;\;}  \;\;\;\;\;\;   \notag \\
   &= \frac{ \log(\sum_{j=1}^m  n_j^{\beta_j})\; - \log(m) +log(\mu) \; }{ \;\;\;\log N \;\;}  \;\;\;\;\;\; (\text{for} \;\;\; N \neq 1), 
\end{align}
where $\overline{\nu}=m/ \mu$, and $\mu$ represents the species turnover rate between sites (see Appendix A).

When $b_j$ and $n_j$ are the same in each site $j$ we can simplify further,
\begin{align}
   {{ \beta^*_{\mathcal{M}}} }  &= \frac{\beta \log n \;+ \log \mu \; }{ \;\;\;\log n \;+ \log \mu \;}  \;\;\;\;\;\; (\text{for} \;\;\; n\times \mu \neq 1) 
\end{align}

\subsubsection{True regional yields based on local exponents} \label{true_regional_yield}
As we pointed out in the main paper, the exponent ${{ \beta^*_{\mathcal{M}}} }$ is not real or meaningful  \emph{saturation} exponent -- it does not actually indicate how some property will change with increasing diversity, unlike the local $\beta$ exponents in our simple set-up. To illustrate this let us first see how the real or \emph{true} ecosystem effects change with increasing biodiversity. We do this by taking a continuously differentiable version of the RHS of Eq. \eqref{eq:Appendix_regional_biomass} with respect to local biodiversity $n$, and take the derivative $ \frac{\mathrm{d}}{\mathrm{d}N} \sum_j y_{_{\mathrm{Total},j} }$ (at a given metacommunity size $m$):
\begin{align}\label{eq:Appendix_ddN_sum_ecosys_a}
  \frac{\mathrm{d}N}{\mathrm{d}n} \cdot \frac{\mathrm{d}}{\mathrm{d}N} \sum_j y_{_{\mathrm{Total},j} } \; &= \; \frac{\mathrm{d}}{\mathrm{d}n} \left(Y \; m\;n^\beta \right)  \notag \\
   \Longrightarrow \quad\quad \frac{\mathrm{d}}{\mathrm{d}N} \sum_j y_{_{\mathrm{Total},j} } \; &= \; \frac{\beta \; m \; Y n^{\beta-1} }{ \frac{\mathrm{d}N}{\mathrm{d}n} }. 
\end{align}
At this point we can note that both $N=m \times \overline{n}$ and $n=\mu \times  \overline{\nu}$ for a given $m$ (see Appendix \ref{saturationexponent}). In the simple case we consider here the diversity levels in all sites will be assumed equal, $n_j= n$ (for all $j$), and all species occupy the same number of sites, $\overline{\nu}=\nu$, which will result in $\mu$ being constant for a given $m$ even as the total  diversity $N$ changes. This means that $N = \frac{m}{\nu} \; n$. Thus the \emph{true} change in an ecosystem property, like yield, that will result from an  increase in the regional biodiversity $N$ will be
\begin{align}\label{eq:Appendix_ddN_sum_ecosys_b}
    \frac{\mathrm{d}}{\mathrm{d}N} \sum_j y_{_{\mathrm{Total},j} }   &=  \nu \; \beta \; Y n^{\beta-1}. 
\end{align}
In the main paper it was assumed that $\nu=1$, and as a result the true regional yield increase was given by the expression $\beta \; Y n^{\beta-1}$ in Eq. \eqref{eq:der_y_sum}.

\end{document}